\documentclass{iopart}
\usepackage{fullpage}
\usepackage{iopams,graphicx,color}

\usepackage{setstack}
\usepackage{upgreek}
\usepackage{array}
\usepackage{ifthen}
\usepackage[pdftex,breaklinks=true,bookmarks=false]{hyperref}

\usepackage{tikz}
\usetikzlibrary{arrows}
\usetikzlibrary[shapes.geometric]
\usetikzlibrary{decorations}

\newcommand{\journ}[5]
{\ifthenelse{\equal{#1}{jsm}}{
{\it #5}, #4 J. Stat. Mech. \href{http://iopscience.iop.org/1742-5468/#4/#2/#3}{#3}}
{\ifthenelse{\equal{#1}{pr}}{
{\it #5}, #4 Phys. Rev {\bf #2} \href{http://link.aps.org/abstract/PR/v#2/e#3}{#3}}
{\ifthenelse{\equal{#1}{prl}}{
{\it #5}, #4 Phys. Rev. Lett {\bf #2} \href{http://link.aps.org/abstract/PRL/v#2/e#3}{#3}}
{\ifthenelse{\equal{#1}{prb}}{
{\it #5}, #4 Phys. Rev. B {\bf #2} \href{http://link.aps.org/abstract/PRB/v#2/e#3}{#3}}
{\ifthenelse{\equal{#1}{pra}}{
{\it #5}, #4 Phys. Rev. A {\bf #2} \href{http://link.aps.org/abstract/PRA/v#2/e#3}{#3}}
{\ifthenelse{\equal{#1}{arxiv}}{
{\it #5}, #4 \href{http://arxiv.org/abs/#2.#3}{arXiv:#2.#3}}
{\ifthenelse{\equal{#1}{rmp}}{
{\it #5}, #4 Rev. Mod. Phys {\bf #2} \href{http://link.aps.org/abstract/RMP/v#2/e#3}{#3}}
{\ifthenelse{\equal{#1}{cond-mat}}{preprint
\href{http://arxiv.org/abs/cond-mat/#2}{cond-mat/#2}}
{\ifthenelse{\equal{#1}{pre}}{
{\it #5}, #4 Phys. Rev. E {\bf #2} \href{http://link.aps.org/abstract/PRE/v#2/e#3}{#3}}
{\it #5}, #4 #1 {\bf #2} #3}}}}}}}}
}

\newcommand{\journdoi}[6]{{\it #5}, #4 #1 {\bf #2} \href{http://dx.doi.org/#6}{#3}}

\begin{document}

\newcommand{\ket}[1]{
	\left| #1 \right>
}

\newcommand{\bra}[1]{
	\left< #1 \right|
}

\newcommand{\Trace}{
	{\rm Tr~}
}

\newcommand{\arctanh}{
	{\rm arctanh}
}
\newcommand{\arccoth}{
	{\rm arccoth}
}

\newcommand{\hh}{\mathbb{H}}

\title[Local quantum quenches in critical one-dimensional systems]
{Local quantum quenches in critical one-dimensional systems: entanglement,
the Loschmidt echo, and light-cone effects}
\author{Jean-Marie St\'ephan$^{1}$ and J\'er\^ome Dubail$^{2}$}

\address{$^1$ Institut de Physique Th\'eorique, CEA, IPhT, CNRS, URA 2306, F-91191 Gif-sur-Yvette, France
\\$^2$ Department of Physics, Yale University, P.O. Box 208120, New Haven, CT 06520-8120, USA}

\eads{\mailto{jean-marie.stephan@cea.fr},\mailto{jerome.dubail@yale.edu}}

%%%%%%%%%%%%%%%%%%%%%%%%%%%%%%%%%%%%%%%%%%%%%%%%%
\begin{abstract}
%%%%%%%%%%%%%%%%%%%%%%%%%%%%%%%%%%%%%%%%%%%%%%%%%
We study a particular type of local quench in a generic quantum critical one-dimensional system, using
conformal field theory (CFT) techniques, and providing numerical checks of the results in free fermion systems.
The system is initially cut into two subsystems $A$ and $B$
which are glued together at time $t=0$. We study the entanglement entropy (EE) between the two parts $A$
and $B$, using previous results by Calabrese and Cardy \cite{CalabreseCardy2004,CalabreseCardy2007},
 and further extending them. We also study in detail the (logarithmic) Loschmidt echo (LLE). For finite-size systems both quantities turn out to be
 (almost) periodic in the scaling limit, and exhibit striking light-cone effects. While these two quantities behave similarly immediately after the quench---namely as $c/3 \log t$ for the EE and $c/4 \log t$ for the LLE---, we
observe some discrepancy once the excitations emitted by the quench bounce on the boundary and
evolve within the same subsystem $A$ (or $B$). The decay of the EE is then non-universal,
as noticed by Eisler and Peschel \cite{EislerPeschel2007}. In contrast, we find that the evolution of the LLE is less sensitive than the EE to non-universal details of the model, and is
still accurately described by our CFT prediction. To further probe these light-cone effects, we also introduce a variant of the Loschmidt echo specifically constructed 
to detect the excitations emitted just after the quench.

%%%%%%%%%%%%%%%%%%%%%%%%%%%%%%%%%%%%%%%%%%%%%%%%%
\end{abstract}
%%%%%%%%%%%%%%%%%%%%%%%%%%%%%%%%%%%%%%%%%%%%%%%%%

\date{\today}
%\submitto{JSM}
\maketitle

\tableofcontents

%%%%%%%%%%%%%%%%%%%%%%%%%%%%%%%%%%%%%%%%%%%%%%%%%
%%%%%%%%%%%%%%%%%%%%%%%%%%%%%%%%%%%%%%%%%%%%%%%%%
\section{Introduction}
\label{sec:intro}
%%%%%%%%%%%%%%%%%%%%%%%%%%%%%%%%%%%%%%%%%%%%%%%%%
%%%%%%%%%%%%%%%%%%%%%%%%%%%%%%%%%%%%%%%%%%%%%%%%%
The characterization of entanglement in quantum many-body systems
has become increasingly important over the last few years. It has been at
the heart of a fruitful interdisciplinary work, with strong motivations
coming both from quantum information \cite{Amico} and from ``conventional''
condensed matter theory. These include topics as diverse as 
the investigation of computational complexity of 
algorithms such as DMRG \cite{White,Schollwock}, or the development of new numerical tools such as PEPS \cite{PEPS} or MERA \cite{MERA}---which
aim to give access to far larger
system sizes than exact diagonalisation techniques---, or the hunt for theoretical tools
able to grasp some of the features of topologically ordered phases of matter \cite{KitaevPreskill, LevinWen, HaldaneLi}.

Among all the competing quantities that have been proposed as (theoretical)
measures of quantum entanglement, the von Neumann entanglement entropy (EE) is the most
widely accepted\cite{Amico,eereview2,eereview3}. It is defined through a bipartition of the Hilbert space
$\mathcal{H} = \mathcal{H}_A \otimes \mathcal{H}_B$.
Most of the time, the bipartition corresponds to a spatial
cut of the system, the subsystem $A$ being thought of as the physical system of
interest, and $B$ as the ``environment''.
Usually, one starts with a pure state $\ket{\Omega} \in \mathcal{H}$,
then defines its reduced density matrix $\rho_A = {\rm Tr}_{\mathcal{H}_B} \ket{\Omega}\bra{\Omega}$, and its EE as the von Neumann entropy of $\rho_A$
\begin{equation}
	S \, =\, -{\rm Tr}_{\mathcal{H}_A} \rho_A \log \rho_A.
\end{equation}
Intuitively, the more entangled the system is, the less separable it is, and therefore 
the larger the EE is. While the EE has been computed numerically or analytically in a
large variety of systems, its experimental
measure would generically require the knowledge of a prohibitively large 
number of correlation functions, which can be traced back to the intrinsic non-locality
of this quantity. However, some methods have been proposed for direct measures in free fermion
systems\cite{KlichLevitov2009,Song2010,SongFlindt2010} (see also the discussion in \cite{HsuMeter} about interacting systems), or for an original experimental implementation of the replica trick which could
possibly lead to a measure\cite{Cardyexp2011} of the R\'enyi entropies $S_n = \frac{1}{1-n}\log {\rm Tr} \rho_A^n$.

In \cite{DS} we introduced a quantity which has a behaviour that is similar
to the EE, the (logarithmic) bipartite fidelity (LBF). The LBF is in principle simpler
to measure than the EE, because it is defined as an overlap of two states (see below).
The common feature of the ``experimental'' methods that have been proposed to measure both the EE
and the LBF is that they rely on probing
some of the non-equilibrium properties of the system. More specifically, the relation
between entanglement and time-evolution of quantum systems after a sudden change
of parameters in the Hamiltonian, {\it i.e.} after a {\it quantum quench}, has been emphasized.
 In that sense, the methods proposed for a measure
of the R\'enyi entropy \cite{Cardyexp2011} and for the LBF \cite{DS} are the same, and rely on the probability of observing the system in the ground state of the Hamiltonian $H'$ after a sudden local quench $H \rightarrow H'$. An experimental device would be considerably more involved in the case of the R\'enyi entropies
\cite{Cardyexp2011} than for the LBF though. The theoretical study of quantum quenches has been very active \cite{Polkovnikov2005,CCquench,BostonGuys,deGrandi2010,deGrandisineGordon,PolkovnikovCo,CardyCalabreseP06008,Spyros2008,MussardoQuench} (the list is far from exhaustive),
in part triggered by the recent experimental achievements in ultracold atoms \cite{Greiner,BlochDalibard}, which now
allow to build atomic systems without dissipation, and with large ranges of tunable
parameters. Another fascinating question is the one of thermalisation in quantum
systems after quenches \cite{Srednicki,Rigol,Kollath,Zanardi,Calabrese}.

One of the most interesting aspect of some of the results on entanglement and quantum quenches
is their universality \cite{Holzhey1994,Vidal2003,Korepin2004}.
Various scaling laws, critical exponents and universal functions have been discovered. 
In the case of one-dimensional systems, in particular, it is well-known that, at quantum
criticality, the low-energy excitations are generically described by
a conformal field theory (CFT). The CFT formalism has been successfully applied to derive
a variety of results in that context (see for example \cite{CalabreseCardy2004,CalabreseCardy2007,DS,Spyros2008} and references therein).

In this paper, we fill some of the gaps that still appear in this already large literature.
Indeed, to our knowledge, the question of finite-size systems and quantum recursion
at criticality has not been addressed in general (see, however, \cite{Igloi} for partial results).
Neither has been studied the LBF after a local quench in
such systems, apart from \cite{DS}. Therefore
we focus on local quantum quenches in finite-size systems, which we study by CFT methods,
extending the results of the pioneering papers by Calabrese and Cardy \cite{CalabreseCardy2007}
and Eisler and Peschel \cite{EislerPeschel2007,EislerPeschel2008} and providing numerical checks of our
results in free fermion systems. We study both the EE and the LBF,
emphasizing the differences between the behaviours of these two quantities. We find
that there are striking light-cone effects, both for the EE and for the LBF.
 Since for a system evolving in time the LBF is nothing but (minus the logarithm of) the {\it Loschmidt echo}
\begin{equation}
 \mathcal{L}(t)=\left| \left< \psi(t) | \psi(0) \right> \right|^2,
\end{equation}
 as noted in \cite{DS}, in the rest of this paper we will prefer the latter terminology, as it is more often used in the context of
quantum quenches. We will therefore call {\it logarithmic Loschmidt echo} (LLE) the
quantity $\mathcal{F}(t) \equiv - \log \mathcal{L}(t)$, keeping in mind that it is the same as the LBF defined in
\cite{DS} evolving with time $t$.

The paper is organized as follows. In Sec.~\ref{sec:finite_size} we first describe some basic consequences of
 a local quantum quench in a one-dimensional critical system. In particular, we explain why the time evolution of
 the wave function after the quench should be periodic. In Sec.~\ref{sec:cft} we derive several results for the time-evolution
 of the entanglement entropy as well as the Loschmidt echo. This section relies mainly on CFT methods, and may be skipped 
at first reading. Finally, Sec.~\ref{sec:plateaus} focuses on the results, providing physical interpretations and
 numerical checks of these in free fermion systems.

%%%%%%%%%%%%%%%%%%%%%%%%%%%%%%%%%%%%%%%%%%%%%%%%%
%%%%%%%%%%%%%%%%%%%%%%%%%%%%%%%%%%%%%%%%%%%%%%%%%
\section{Finite-size critical systems in one dimension}
\label{sec:finite_size}
%%%%%%%%%%%%%%%%%%%%%%%%%%%%%%%%%%%%%%%%%%%%%%%%%
%%%%%%%%%%%%%%%%%%%%%%%%%%%%%%%%%%%%%%%%%%%%%%%%%
In this paper we consider 1+1-dimensional systems which are described
by a CFT at sufficiently large distances and time scales (see for example \cite{CFT1,CFT2} for an introduction to CFT). 
We are interested mainly in finite-size systems.
 Typically, such a system can be a gapless spin chain of finite size $L$, for sufficiently
large $L$. For a system with open boundary conditions, which are assumed
to correspond to a single conformal boundary condition throughout the paper,
the lower-energy part of the spectrum of the gapless Hamiltonian
$H_L$ is given by \cite{CBNc,Affleckc}
\begin{equation}
	\label{eq:finitesizec}
	E_{\alpha} \, = \, L \,E_{{\rm bulk}} + \, E_{{\rm boundary}} \,+\,  \frac{\pi \, v_F }{L} \left(h_\alpha - \frac{c}{24} \right) \, + \, O\left(1/L^2\right)
\end{equation}
$L\, E_{{\rm bulk}}$ is the (bulk) extensive part of the energy, and $E_{{\rm} boundary}$ the boundary energy.
 $E_{{\rm bulk}}$ and $E_{{\rm} boundary}$ are both non-universal.
$v_F$ is the Fermi velocity, $c$ is the (universal) central charge of the CFT
and the $h_\alpha$ are the conformal dimensions of the
operators $\phi_\alpha$ appearing in the low-energy part of the spectrum.

There are two kinds of such operators, the {\it primary fields} and their {\it descendants}.
 The descendants can always be constructed out of a single primary field,
 and correspond to some combinations of its derivatives. The primary fields cannot in
general be expressed as descendants of other fields. The primary fields that
appear in the spectrum (\ref{eq:finitesizec}) depend both on the CFT and on
the boundary condition we are dealing with. Typically, in unitary CFTs that
are of interest for condensed matter applications (for example as scaling
limits of spin chains), there is a discrete (often finite) set of primary fields.
If a primary field is present in the spectrum (\ref{eq:finitesizec}), then all its descendants must
appear as well. The conformal dimensions of a descendant $\phi_{a,k}$ and of the primary
field $\phi_a$ are {\it always separated by an integer}. This property is crucial
when we deal with time evolution. Indeed, if the system is initially in the state
$\ket{\psi(0)}$, then $\ket{\psi(t)} = e^{-i H t} \ket{\psi(0)}$, and in particular
after time $t=2 L/v_F$ all the descendants of a primary field $\phi_\alpha$ pick
up the same phase, up to additive $O(1/L^2) \times L = O(1/L)$ corrections.

\begin{figure}[ht]
\begin{center}
\includegraphics{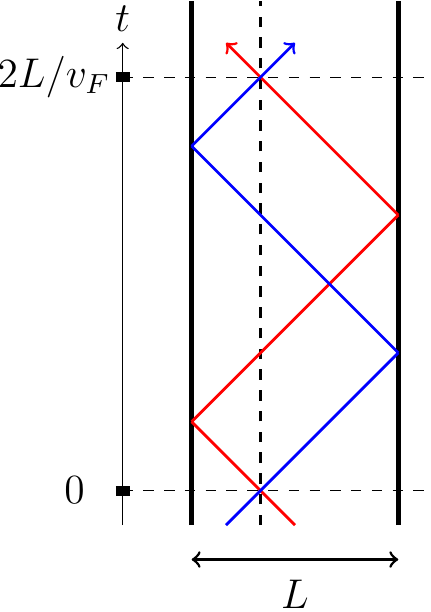}
\end{center}
\caption{The ``quasi-particle'' picture gives a useful though heuristic way of thinking about
the time-evolution of finite-size critical systems in 1d. All the quasi-particles move at speed $v_F$.
After time $2L/v_F$, the particles come back at the same
position. Therefore, up to possible phase factors due to the reflection against the boundaries,
the system is periodic.}
\label{fig:particles}
\end{figure}

\paragraph{}
This can be interpreted intuitively with the following semi-classical picture
(see also Fig. \ref{fig:particles}). Low-energy excitations can be seen as
some ``quasi-particles'' moving at fixed speed $v_F$. There can be different species $a$ of
quasi-particles in the system, which correspond to the different {\it primary} fields $\phi_a$ in the CFT.
 The behaviour of these different species of quasi-particles differs
at the boundary: they pick up a phase factor $e^{-i \pi h_a}$. This ``quasi-particle''
picture is heuristic and is useful to predict light-cone behaviours, as explained below,
however it fails to give quantitative predictions in general.

\begin{center}
\begin{figure}[ht]
\includegraphics[width=8cm]{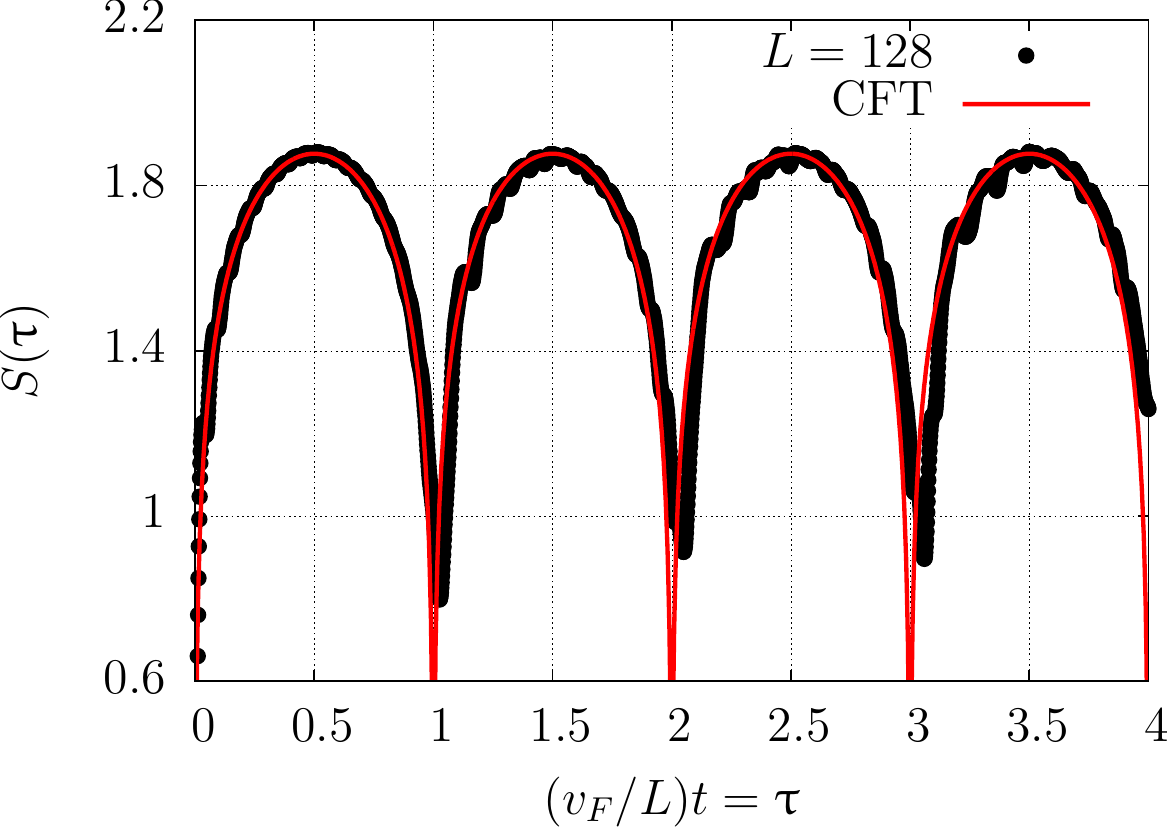} \quad
\includegraphics[width=8cm]{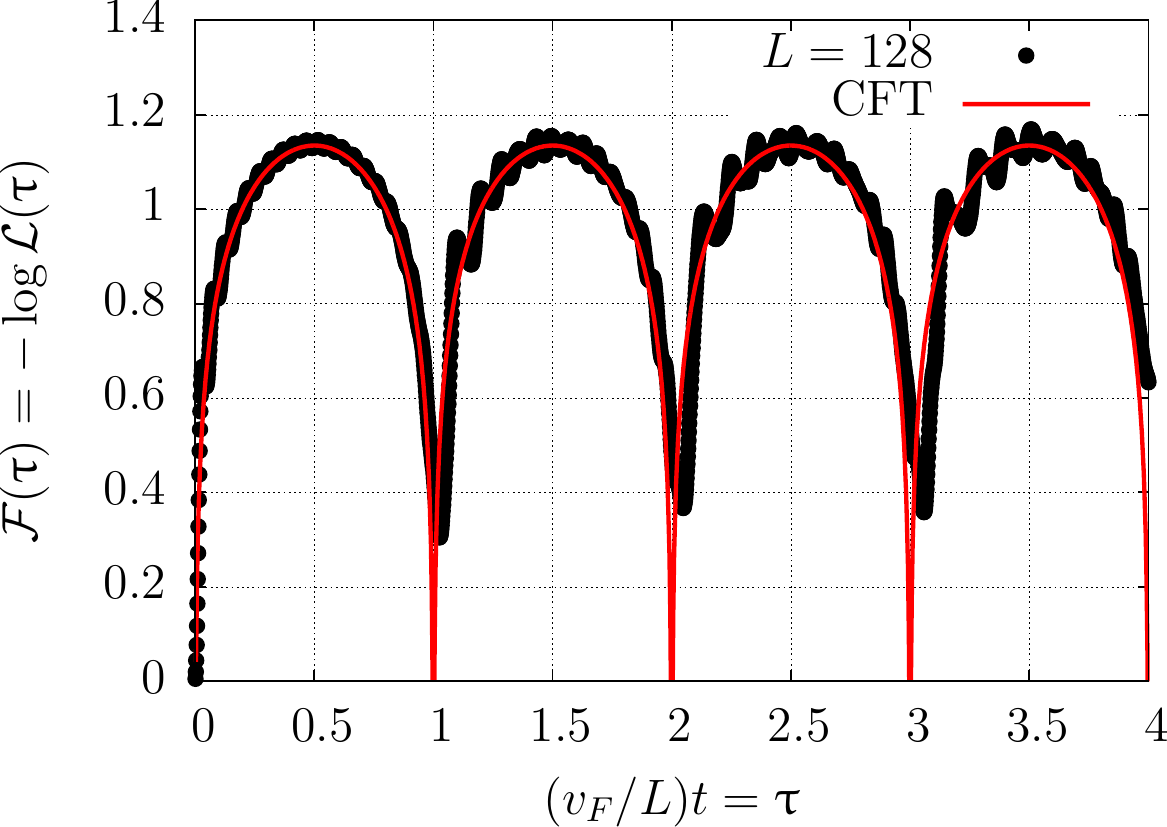}
\caption{Entanglement entropy and the logarithmic Loschmidt Echo against rescaled time $(v_F/L)t$ after a local quench in the middle of a system
of size $L=128$. A couple of oscillations of the finite-size system can be observed after the local quench.
 They are very well described by the CFT (slightly anticipating on Eq.~\ref{eq:le_sym} and Eq.~\ref{eq:s_sym}).
 The rapid oscillations appearing in the plots are analogous to the one observed at equilibrium
 (see for example \cite{ScalingCorrections1,ScalingCorrections2,ScalingCorrections3}). Their amplitude decreases for bigger and bigger system sizes,
 and we will not discuss them in this paper. After longer times,
the non-universal $O(1/L^2)$ terms in (\ref{eq:finitesizec}) spoil the CFT oscillations, and the periodic behaviour is lost.}
\label{fig:oscillations}
\end{figure}
\end{center}

\paragraph{}
It is important at this stage to emphasize that the commensurability 
of the spectrum (more precisely of each subsector of the spectrum corresponding to
each primary field $\phi_a$ and its descendants) can only occur in the scaling limit.
It is not true in a finite system, because of the order $O(1/L^2)$ corrections in
(\ref{eq:finitesizec}). The crucial requirement is that the dispersion relation
must be linear $E_{k} = v_F |k|$ (if $k$ is the momentum in a translation-invariant system). In our numerical checks we will focus on
finite-size lattice systems, and although the dispersion relation
can be linearised close to the Fermi energy, there are subleading terms
which spoil the commensurability of the spectrum (see \cite{CardyCalabreseP06008} for a detailed discussion of this point). However, as we shall see,
the periodic behaviour will be a very
good approximation in the case of the local quenches we are interested in: see already Fig.~\ref{fig:oscillations} for a first example. There
are two reasons why this is the case.
\begin{itemize}
	\item We look at the behaviour of the system at sufficiently large
distance and time scales (typically large compared to the lattice spacing $a$),
such that the continuum description is relevant. However, we restrict our analysis
to times $t \ll L^2/(a v_F)$, such that the order $O(1/L^2)$ terms in (\ref{eq:finitesizec})
are still small. For larger times, these terms spoil the periodicity. Typically, for
systems with a number of sites of order $L/a \approx 100$, we expect to see about $10$
oscillations that are well described by the CFT calculations, and after that the non-universal corrections become too large.
	\item We are interested in local quenches only.
 The Hamiltonian $H_L \rightarrow H'_L$	is changed only locally. This is radically different from a {\it global} quench, where
	parameters of the Hamiltonian would be changed suddenly everywhere in the system.
	After a local quench, there are much less excitations in the system than after a
global one. In particular, for a local quench, we expect only the lower energy part of the spectrum to be significantly excited.
 For these excitations the formula (\ref{eq:finitesizec}) applies. For {\it global} quenches however, we would expect also high-energy excitations with
non-universal scaling behaviour. Lattice effects would be presumably much more important in that case,
 and would spoil the commensurability of (the different sectors of) the spectrum almost
immediately, leading to chaotic or turbulent behaviour, as observed in \cite{Zanardi}.
\end{itemize}

%%%%%%%%%%%%%%%%%%%%%%%%%%%%%%%%%%%%%%%%%%%%%%%%%
%%%%%%%%%%%%%%%%%%%%%%%%%%%%%%%%%%%%%%%%%%%%%%%%%
\section{Local quenches from conformal field theory}
%%%%%%%%%%%%%%%%%%%%%%%%%%%%%%%%%%%%%%%%%%%%%%%%%
%%%%%%%%%%%%%%%%%%%%%%%%%%%%%%%%%%%%%%%%%%%%%%%%%
\label{sec:cft}
We now specify the type of local quenches we are interested in, and we show how these
can be studied using CFT techniques, following previous works \cite{CalabreseCardy2007,EislerPeschel2008,DS}.

\begin{figure}[ht]
\begin{center}
\includegraphics{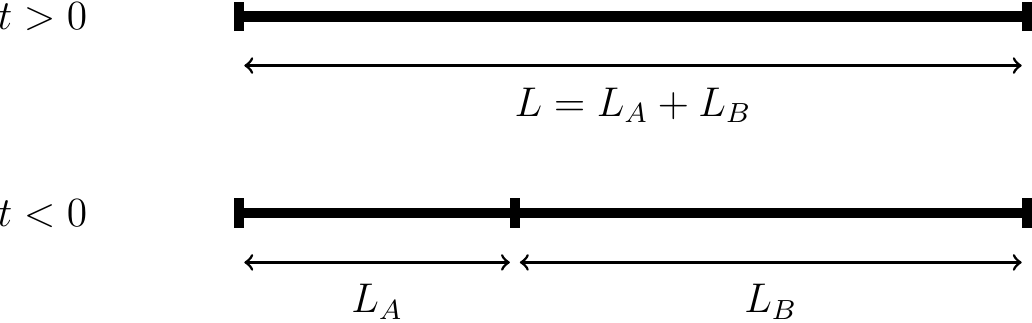}
\end{center}
\caption{The system is initially prepared in the ground state of $H_A+H_B$, then at time $t>0$ it evolves with the total Hamiltonian
 $H = H_A+H_B+H_{AB}^{\rm I}$.}
\label{fig:quenchline}
\end{figure}

\subsection{Cut and glue: the type of local quench}
In this paper we study one type of local quench only. Let us consider
a one-dimensional system of length $L$, with open boundary conditions.
We cut the system into two parts $A$ and $B$, of length $L_A$ and $L_B=L-L_A$
respectively. The system is initially prepared in the ground state of the
Hamiltonian $H_A + H_B$, which is of course the tensor product $\ket{A} \otimes \ket{B}$ of the ground state $\ket{A}$ of $H_A$ and the ground state $\ket{B}$
 of $H_B$. For later convenience we note $\ket{A\otimes B} = \ket{A} \otimes \ket{B}$. At time $t=0$, the
interaction $H_{AB}^{\rm I}$ between $A$ and $B$ is switched on instantaneously. The system then
evolves with the total Hamiltonian $H=H_A + H_B +H_{AB}^{\rm I}$ (see Fig.~\ref{fig:quenchline}).

Throughout this paper we consider situations when the boundary condition (b.c) is always
the same. Namely, if we are dealing with a spin chain of size $L$ with free b.c,
then the boundary conditions at the cut between $A$ and $B$ are also free. Moreover,
we assume that this is a conformal b.c. \cite{CardyCBC}. This is always true at least in the
scaling limit, since every b.c should renormalize towards a conformal b.c
under the renormalization group (RG) flow. In this paper we address neither the question of
different b.c, which would involve boundary condition changing operators, nor the one
of boundary RG flows.

\paragraph{}

There is an important consequence to these assumptions. In general, the initial state
$\ket{A \otimes B}$ should be decomposed as
\begin{equation}
	\ket{A \otimes B} \, = \, \sum_{a} \sum_k c_{a,k}  \ket{\phi_{a,k}}
\end{equation}
where the index $a$ runs over the different primary fields in the spectrum of the CFT,
and $k$ runs over the descendants of each primary field. This, in principle, gives
the time-evolution after the quench, since by definition the $\ket{\phi_{a,k}}$ are
the eigenstates of the Hamiltonian $H$, with eigenvalues (\ref{eq:finitesizec}). The
calculation of the coefficients $c_{a,k}$ is difficult though. {\it For the particular
type of quench we are looking at, however, only the identity and its descendants
have non-zero overlaps} $c_{a,k} = \left< \phi_{a,k} | A \otimes B\right>$. {\it This huge
simplification occurs because we assume that the conformal b.c. is the same everywhere.
} Since the initial state $\ket{A \otimes B}=  \ket{A} \otimes \ket{B}$ is built out
of the ground states of $H_A$ and $H_B$, which are both Hamiltonians with the same b.c.
on the left and right boundary, $\ket{A}$ and $\ket{B}$ both correspond to the identity
operator. Then the fusion of these two states has to be the identity as well. This,
composed with some conformal transformation of the space-time geometry
(as will become clear below) cannot generate any operator but the identity and
its descendants.

Since only the sector of the identity appears in the spectrum, the evolution of the system should be
truly periodic (up to the order $O(1/L^2)$ corrections discussed in the previous section).
 There is no additional phase corresponding to different primary operators which could
presumably affect the periodicity anymore. In the ``quasi-particle'' picture this
means that there is only one species of quasi-particles which can be emitted by the
local quenches we are considering.

Let us stress again the fact that this conclusion would not hold if we were considering different b.c. in the system.
 Changing the b.c., for example on both sides of the quench,
would generate other species of quasi-particles.

\subsection{Space-time geometry and conformal mappings}
As a warm-up exercise, let us consider the simpler case of two semi-infinite systems $A$
and $B$ that are glued together at time $t=0$.
This of course corresponds to the limiting case $a \ll v_F\, t \ll L_A,L_B$ in the above setting (Fig. \ref{fig:quenchline}). We review the CFT calculation
of the Loschmidt echo \cite{DS} and of the entanglement entropy \cite{CalabreseCardy2007}
 in that case.
\subsubsection{Loschmidt echo}
\begin{figure}[ht]
	\begin{center}
	\includegraphics{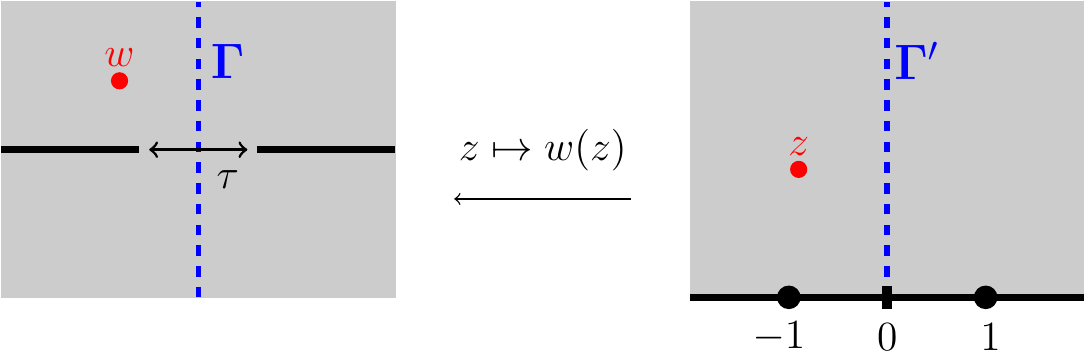}
	\end{center}
	\caption{The geometry used to define the partition function $Z_{A,B}$.
 It can be obtained by a conformal mapping from the upper half-plane $z \mapsto w(z) = (\tau/4)(z+z^{-1})$. Notice that the imaginary time
 flows in the left-right direction. The distance between the
 two slits is $\tau$. The infinitesimal transformation $w\mapsto w(1+\delta \tau/\tau)$ is applied at the right of $\mathbf{\Gamma}$ in the $w$ region.
 $\mathbf{\Gamma'}$ is the image of $\mathbf{\Gamma}$ through the inverse conformal mapping:  $\mathbf{\Gamma'}=w^{-1}(\mathbf{\Gamma})$.}
	\label{fig:corners}
\end{figure}

The CFT calculation is performed in imaginary time $\tau$. For the Loschmidt echo we need
to compute
\begin{equation}
	\label{eq:loschmidt}
	\mathcal{L}(\tau) =  \left| \bra{A \otimes B} e^{- \tau H} \ket{A \otimes B} \right|^2,
\end{equation}
 where $H=H_{A\cup B}$ is the Hamiltonian of the total system. This can be done using the relation
\begin{equation}\label{eq:Z}
 \ket{A \otimes B} = \lim_{\lambda \rightarrow \infty} e^{-\lambda (H_A+H_B)} e^{\lambda E_0} \ket{s} / \left< s|A \otimes B \right>, 
\end{equation}
 which is valid for any
state $\ket{s}$ such that $\left< s|A \otimes B \right> \neq0$.
 $E_0$ is the ground state energy of $H_A+H_B$. The relation (\ref{eq:loschmidt}) becomes (for $\tau >0$)
\begin{equation}
	\label{eq:loschmidtZ}
	\mathcal{L}(\tau) \, =\, {\rm cst} \, \times \, \left| \lim_{\lambda \rightarrow \infty} Z_{A, B}(\tau) \right|^2
\end{equation}
where $Z_{A, B}(\tau) = \left<s \right| e^{-\lambda (H_A+H_B)} e^{-\tau H} e^{-\lambda (H_A+H_B)} \ket{s}$.
 This is the partition function of a statistical mechanical system defined in
the geometry shown in Fig. \ref{fig:corners}. The normalization constant in (\ref{eq:loschmidtZ}) will be fixed later.

There is an extensive part in the free energy $\mathcal{F}_{A,B}(\tau) =  - \log Z_{A,B}(\tau)$ coming from the region
of size $\tau$ between the two corners in Fig. \ref{fig:corners}. This part will contribute
only to a phase factor when we come back to real time ($\tau \rightarrow i t$),
which disappears in the norm (\ref{eq:loschmidt}). Also, there is an extensive free energy
which grows with the size of the system and diverges as $\lambda \rightarrow \infty$. We
can absorb this contribution in the normalization constant however, which will be fixed later
when we come back in real time. Thus, the interesting part in the calculation of the time-evolution of the Loschmidt echo is precisely
 the universal part coming from the corners in Fig.~\ref{fig:corners}. As already emphasized in \cite{DS}, the simplest way
 to get the result at this point is to apply the Cardy-Peschel formula \cite{CardyPeschel}. However, we wish to compute the Loschmidt echo
 in more complicated geometries later on, and for pedagogical purposes we will follow a slightly different route.

The dependence of $\mathcal{F}(\tau)= - \log\mathcal{L}(\tau)$ on the distance $\tau$ in Fig.~\ref{fig:corners} can be evaluated as follows \cite{CardyPeschel}.
 Consider the transformation $w \mapsto w+\delta \tau$ for $w$ at the right of the path $\mathbf{\Gamma}$ in Fig. \ref{fig:corners},
 and $w \mapsto w$ at the left of $\mathbf{\Gamma}$. Notice that this mapping changes the distance $\tau$ into $\tau + \delta \tau$.
 The variation of $\mathcal {F}$ can be expressed in terms of the stress-tensor $T(w)$ (and its anti-holomorphic part $\bar{T}(\bar{w})$) as
\begin{equation}
	\frac{1}{2} \delta \mathcal{F} = \frac{\delta \tau}{2\pi} \int_\mathbf{\Gamma} \left<T(w) \right> dw \; + \; c.c.
\end{equation}
and the one-point function of the stress-tensor can be obtained from a conformal mapping from the half-plane (see Fig. \ref{fig:corners})
\begin{equation}
	\label{eq:mappinginfinite}
w(z)=\frac{\tau}{4}(z+z^{-1}).
\end{equation}
The one-point correlation function of $\left< T(z)\right>$ is zero in the half-plane.
 $\left< T(w)\right>$ can be calculated using the transformation law of the stress-tensor\cite{CFT1,CFT2}
\begin{equation}
 T(w) = \left(dw/dz \right)^{-2} \left[T(z) - S(w,z) \frac{c}{12} \right],
\end{equation}
where $S(w,z)$ is the Schwarzian derivative of the conformal transformation $w(z)$ :
\begin{equation}
S(w,z)=\frac{w'''(z)}{w'(z)}-\frac{3}{2}\left(\frac{w''(z)}{w'(z)}\right)^2. 
\end{equation}
We therefore get
\begin{equation}
	\frac{1}{2} \delta \mathcal{F} = -\frac{c \, \delta \tau}{12 \pi}\int_{\mathbf{\Gamma'}} dz \,S(w,z)\left(\frac{dw}{dz}\right)^{-1}
\end{equation}
where the path $\mathbf{\Gamma'}$ is
the image of $\mathbf{\Gamma}$ under the inverse map $w \mapsto z$. 
For the mapping in Eq.~\ref{eq:mappinginfinite}, the integrand $f(z)$ is
\begin{equation}\label{eq:integrand}
 f(z)=S(w,z)\left(\frac{dw}{dz}\right)^{-1}=-\frac{24}{\tau}\frac{z^2}{(z-1)^3 (z+1)^3}.
\end{equation}
Notice that $f(z)$ has singularities at $z=\pm 1$, which are the images of the two tips of the slits in Fig.~\ref{fig:corners} through
 the inverse mapping $w\mapsto z$. 
In this case we have $\mathbf{\Gamma}=i\mathbb{R}$ and $\mathbf{\Gamma'}=i\mathbb{R}_+$. The integral can be evaluated as follows.
 First we consider both $\mathbf{\Gamma'} = i \mathbb{R}_+$ and its mirror image: $i \mathbb{R}_+ \to i\mathbb{R}$, which multiplies the integral by $2$. Then we close
 the contour by a big half-circle to the right. We call this contour $\mathcal{C}$, then
\begin{equation}
 \int_{\mathbf{\Gamma'}}f(z)\,dz=\frac{1}{2}\int_{i\mathbb{R}}f(z)\,dz=\frac{1}{2}\int_{\mathcal{C}}f(z)\,dz.
\end{equation}
Finally the integral over $\mathcal{C}$ can be computed using the residue theorem. Only the pole at $z=1$ will contribute, and we obtain
\begin{equation}\label{eq:all}
\boxed{
 \frac{\delta \mathcal{F}}{\delta \tau}=\frac{c}{6}\times {\rm Res} \big[f(z)\,;\,z=1\big]
}
\end{equation}
This equation is particularly useful because the computation of the Loschmidt echo for all the other geometries will lead to a similar result.
 Namely, the calculation boils down to a residue at the image of one of the tips through the inverse mapping. Here we get 
$
\delta \mathcal{F}/\delta \tau = \frac{c}{4\tau}$, which gives finally
\begin{equation}
 \mathcal{F}(\tau) = -\log \mathcal{L}(\tau) = \frac{c}{4} \log |\tau| +\; {\rm cst}.
\end{equation}
Now we can perform the Wick rotation $\tau \rightarrow i v_F t +\epsilon$ and come back in real time. 
An UV cutoff $\epsilon$ is needed here in order for the Loschmidt echo to be defined at time $t=0$ (note that $\mathcal{L}(t=0)=1$). 
It is of the order of a lattice spacing in the case of spin chains. The logarithmic Loschmidt echo (LLE) is then
\begin{equation}
 \mathcal{F}(t) \, =\, -\log \mathcal{L}(t) \,=\, \frac{c}{4} \log \sqrt{1+ (v_F t) ^2/\epsilon^2} \, \simeq\, \frac{c}{4} \log t + {\rm cst}
\end{equation}
when $t \gg \epsilon$, which is the result of \cite{DS}.
\subsubsection{Entanglement entropy}
We recall quickly the calculation of the entanglement entropy, following Calabrese and Cardy\cite{CalabreseCardy2007}. The R\'enyi entropy
\begin{equation}
 S_n=\frac{1}{1-n} \log \left({\rm Tr}\,\rho^n\right),
\end{equation}
gives the EE in the limit $n \rightarrow 1$, assuming that $S_n$ is analytic in $n$
\begin{equation}
 S=-{\rm Tr }\;\rho \log \rho= \lim_{n\to 1} S_n
\end{equation}
When $n\geq 2$ is an integer, ${\rm Tr} \,\rho^n$ admits a simple interpretation in terms of partition functions on a $n-$sheeted Riemann surface,
 and can be computed in CFT\cite{CalabreseCardy2004,Holzhey1994}. While the validity of the replica trick here is by no means obvious,
 the method has enjoyed considerable success so far and has proved to be very powerful in the context of 1D quantum critical systems. 
\paragraph{}
As explained in detail in \cite{CalabreseCardy2004,CalabreseCardy2007}, ${\rm Tr}\, \rho^n$ transforms as a one-point function of a
 primary field $\Phi_n$ under a conformal transformation. Its conformal dimension is
\begin{equation}
 x_n=\frac{c}{12}\left(n-\frac{1}{n}\right)
\end{equation}
\begin{figure}[ht]
	\begin{center}
	\includegraphics{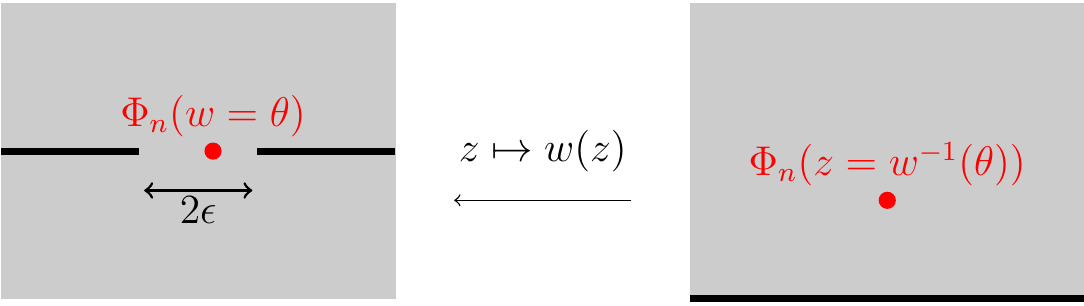}
	\end{center}
	\caption{The geometry used to compute the entanglement entropy. $\Phi_n$ is the twist operator
introduced by Calabrese and Cardy in \cite{CalabreseCardy2004}.}
	\label{fig:corners_ee}
\end{figure}
The conformal mapping needed is the same as for the Loschmidt echo, with $\tau$ changed into $2 \epsilon$ (see Fig.~\ref{fig:corners_ee})
\begin{equation}\label{eq:mapping_ee_inf}
 w(z)=\frac{\epsilon}{2}(z+z^{-1}).
\end{equation}
The R\'enyi entropy is given in this CFT calculation by
\begin{equation}
 S_n \sim \frac{1}{1-n}\log \big|\langle \Phi_n (w) \rangle\big|,
\end{equation}
The one point function can be obtained by mapping back to the $z-$half plane (Fig.~\ref{fig:corners_ee})
\begin{equation}
\langle \Phi_n(w) \rangle\propto \left| \frac{1}{\Im {\rm m}(z)}\frac{dz}{dw}\right|^{x_n},
\end{equation}
and the entropy is expressed as
\begin{equation}
 S_n\sim -\frac{c}{12}\left(1+\frac{1}{n}\right)\log \left|\frac{1}{\Im{\rm m}(z)}\frac{dz}{dw}\right|
\end{equation}
Since the initial cut coincides with the bipartition, the operator lies
 between the two slits at some point $w=\theta$ on the real
axis, with $-\epsilon<\theta<\epsilon$ (see Fig.~\ref{fig:corners_ee}). Inverting (\ref{eq:mapping_ee_inf}), we get 
\begin{equation}
 \left|\frac{1}{\Im{\rm m}(z)}\frac{dz}{dw}\right|=\frac{\epsilon}{\epsilon^2-\theta^2}
\end{equation}
 Performing the Wick rotation $\theta \to i v_F t$ we recover the celebrated result \cite{CalabreseCardy2007} for the EE $(n=1)$
\begin{equation}
 S(t) \, =\, \frac{c}{3} \log t + {\rm cst}
\end{equation}
in the limit $t \gg \epsilon$.

\subsection{A simple finite size example}
\subsubsection{Loschmidt echo}

\begin{figure}[ht]
\begin{center}
\includegraphics{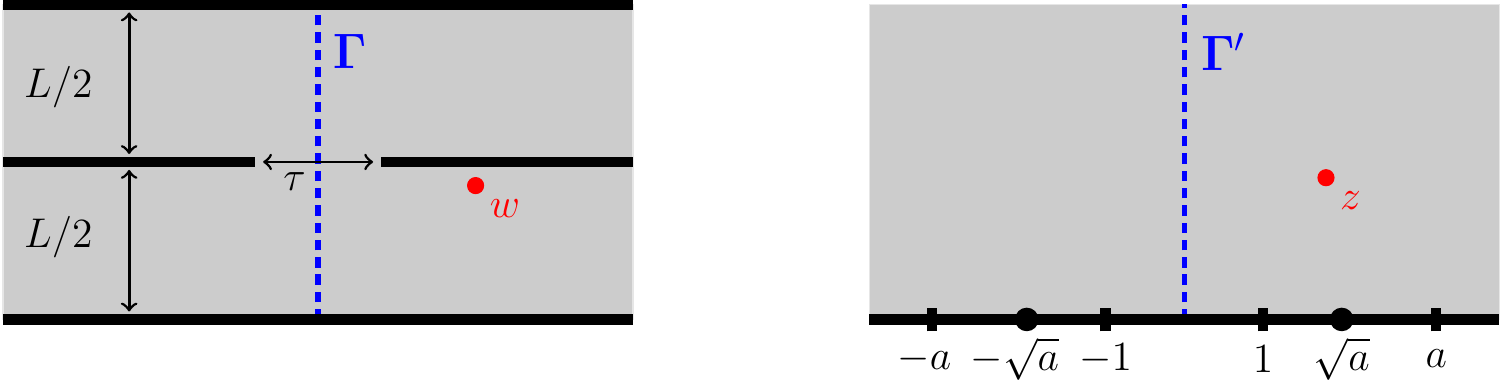}
\end{center}
\caption{Pants geometry (left) and upper half-plane $\mathbb{H}$ (right). The conformal transformation
$w(z)$ is given by Eq.~\ref{eq:conf_pants}.}
\label{fig:pants}
\end{figure}

We consider now the case of a cut in the middle of the chain $L_A=L_B=L/2$. The conformal transformation
 which maps the upper-half plane onto the ``double-pants'' geometry in Fig.~\ref{fig:pants} is 
\begin{equation}\label{eq:conf_pants}
 w(z)=\frac{L}{2\pi} \left[\log \left(\frac{z-1}{z+1}\right)+\log \left(\frac{z-a}{z+a}\right)  - i \pi \right]
\end{equation}
This Schwarz-Christoffel mapping has singularities at $z=\pm 1,z=\pm a$, and extrema at
$z=\pm \sqrt{a}$. The distance between the two slits is 
\begin{equation}\label{eq:tau_pants}
 \tau= \left| w(\sqrt{a})-w(-\sqrt{a}) \right| =\frac{4L}{\pi}\arctanh \left(1/\sqrt{a}\right)
\end{equation}
Inverting this relation, we get
\begin{equation}
 a(\tau)=\coth^2 \left(\frac{\pi \tau}{4L}\right)
\end{equation}
Denoting as before $f(z)=S(w,z)(dw/dz)^{-1}$, the residue we have to calculate is then
\begin{equation}
 {\rm Res}[f(z);z=\sqrt{a}]=\frac{3 \pi}{8L} \frac{a^2+6a+1}{\sqrt{a}(a+1)},
\end{equation}
and the variation of the (logarithmic) Loschmidt echo with the imaginary time $\tau$ is
\begin{equation}
 \frac{\delta \mathcal{F}}{\delta \tau} = -\frac{\delta \log \mathcal{L}}{\delta \tau}=\frac{\pi c}{16 L} \frac{a^2+6a+1}{\sqrt{a}(a+1)}=\frac{\pi c}{4 L}\frac{1}{\tanh (\pi \tau/L)}
\end{equation}
Integrating over $\tau$, we get
\begin{equation}\label{eq:loschmidt_s}
 \mathcal{F}(\tau)= -\log \mathcal{L}(\tau)=\frac{c}{4} \log \left| \sinh (\pi\tau/L) \right|+{\rm cst}.
\end{equation}
Of course, the constant in (\ref{eq:loschmidt_s}) can still have a non-trivial dependence in $L$.
 This is indeed the case as can be shown using once again the Cardy-Peschel formula\cite{CardyPeschel}, with two angles $2\pi$. \footnote{
One can also use the method exposed here using an infinitesimal transformation $L \to L+\delta L$.} This yields another 
logarithmic contribution $(c/4) \log L$. Performing the analytic continuation
 $\tau \to i v_F t+\epsilon$, the result can be written as
\begin{equation}\label{eq:le_sym}
 \mathcal{F}(t) \simeq \frac{c}{4} \log \left|\frac{L}{\pi} \sin \frac{\pi v_F t}{L}\right|+{\rm cst} \qquad (t \gg \epsilon). 
\end{equation}
\subsubsection{Entanglement entropy}
We will here compute the EE for the case of an initial cut in the middle of the chain $L_A=L_B=L/2$. The bipartition coincides with the cut
 for now. As emphasized before\cite{CalabreseCardy2007}, the calculation of the EE requires the analytic inversion of the conformal mapping
 (Eqs.~\ref{eq:conf_pants} and \ref{eq:tau_pants} with $\tau \rightarrow 2 \epsilon$ as in the previous section and the same
 goes for Fig.~\ref{fig:pants}).
 This is important in order for the final analytic continuation
$\theta \rightarrow iv_F t$ to be meaningful. In this simple geometry it is indeed possible:
\begin{equation}
 z(w)= \coth \left(\frac{\pi \epsilon}{2L}\right) \frac{1+\zeta}{1-\zeta}  \qquad {\rm with} \qquad \zeta = \sqrt{\frac{\sinh \frac{\pi}{L}(w+\epsilon)}{\sinh \frac{\pi}{L}(w-\epsilon)}}
\end{equation}
The one point function of the Calabrese-Cardy twist operator must be evaluated at some point
$w=\theta$, with $\theta$ real and $-\epsilon<\theta<\epsilon$
\begin{eqnarray}
 \langle \Phi_n(w) \rangle &\propto& \left|\frac{1}{\Im{\rm m}(z)}\frac{dz}{dw}\right|^{x_n}\\
&\propto&\left|
\frac{\pi \sinh \left[\frac{\pi}{L} 2\epsilon\right]}{2\sinh\left[\frac{\pi}{L}(\epsilon+\theta)\right]\sinh\left[\frac{\pi}{L}(\epsilon-\theta)\right]}\right|^{x_n}
, 
\end{eqnarray}
Taking the analytic continuation $\theta\rightarrow i v_F t$, we find that
in the limit $v_F t \gg \epsilon$
\begin{equation}\label{eq:s_sym}
 S(t)=\frac{c}{3} \log \left|\frac{L}{\pi} \sin \frac{\pi v_F t}{L}\right|+{\rm cst}
\end{equation}
This precise result had already been guessed from numerical simulations in the quantum Ising chain in transverse field, see Ref.~\cite{Igloi,IgloiNew}.
It is also possible to treat the more general case when the bipartition ({\it i.e.} the position of the twist operator)
 does not coincide with the position of the quench, as in \cite{CalabreseCardy2007} for the finite case. In our setup the quench still happens
 at $L/2$, but the EE is measured for a subsystem $A$ of length $L_A=L/2-l$. The twist operator must therefore be evaluated at
 $w=\theta+il$, with $\epsilon<\theta<\epsilon$. The calculation is more cumbersome, but can nevertheless be carried to the
 end. We find after the analytic continuation $\theta \to it$, setting $v_F=1$ for convenience.
\begin{eqnarray}\fl
 S(t)\sim\frac{c}{12}\log \left|\textstyle{
\left(\frac{L}{\pi}\right)^2
\left(\cos^2 \frac{\pi(t+i\epsilon)}{L}-\cos^2 \frac{\pi l}{L}
\right)
\left(
\cos \frac{2\pi l}{L} \cosh \frac{2\pi \epsilon}{L}-\cos\frac{2\pi t}{L}+
2\left|\cos^2 \frac{\pi(t-i\epsilon)}{L}-\cos^2 \frac{\pi l}{L}\right|
\right)
}
\right|
\end{eqnarray}
For $t\gg \epsilon$ this expression can be further simplified (we reintroduce $v_F$ in the formulae). 
\begin{equation}\fl
 S(t)=\left\{
\begin{array}{lll}
 \frac{c}{6} \log L+{\rm cst} &,& 0<v_F t<l\;{\rm or}\;L-l<v_Ft<L\\\\	
\frac{c}{6} \log \left\{\left(\frac{L}{\pi}\right)^2\left[\sin^2 \left(\frac{\pi v_F t}{L}\right)-\sin^2 \left(\frac{\pi l}{L}\right)\right]\right\}+{\rm cst}'
&,&l<v_Ft<L-l
\end{array}
\right.
\end{equation}
$S(t)$ is of course periodic with period $L/v_F$. Notice also that the EE is constant at the beginning,
 until time $l/v_F$.
 A limit of interest is that of an infinite system $l,t \ll L$:
\begin{equation}
 S(t)=\left\{
\begin{array}{lll}
 \frac{c}{6}\log l+{\rm cst}&,&t<l\\\\
\frac{c}{6}\log \left(t^2-l^2\right)+{\rm cst}'&,& t>l
\end{array}
\right.
\end{equation}
where we recover the Calabrese-Cardy\cite{CalabreseCardy2007} result. Actually, one finds
here that our results are nothing but the Calabrese-Cardy results in an infinite geometry,
where $v_F t$ and $l$ have been replaced by their ``chord counterparts'' $\frac{L}{\pi}\sin \left(\frac{\pi v_F t}{L} \right)$
 and $\frac{L}{\pi}\sin \left(\frac{\pi l}{L}\right)$. 
This simple relation with the infinite case is standard in CFT\cite{CalabreseCardy2007}, but breaks down as soon as the quench is not located
exactly in the middle of the strip (like in Fig.~\ref{fig:asymmetric_pants} if $L_A=\ell \neq L/2$).
 See Eqs.~\ref{eq:ee_ns_1s3_a} and \ref{eq:ee_ns_1s3_b} for an example where this trick does not apply (with $L_A=L/3$, $L_B=2L/3$).
%In that case (and only in
%that case), the pants geometry (Fig.~\ref{eq:s_sym}) can be folded along the axis containing the two slits, and viewed as a double-layer system on a strip of width $L/2$
%with two boundary condition changing operators inserted on the folding axis. The same
%trick, applied to the infinite geometry studied by Calabrese and Cardy, gives a double-layer
%system on the half-plane. The two systems are conformally equivalent, and this is why it is
%sufficient to replace $v_F t$ by $\sin \left(\frac{\pi v_F t}{L} \right)$ in the
%Calabrese-Cardy formulae to get the correct result for the strip with a quench in the middle.

\subsection{Semi-analytic method for the Loschmidt echo in other geometries}
\label{sec:cft_ns_infinite}
We are now interested in geometries where the cut is not in the middle of the chain ($L_A \neq L_B$).
 Usually, the loss of the left-right (when $L_A=L_B$) symmetry makes the problem much more complicated, and a full explicit solution for the EE and the LLE becomes less likely. However, we wish to show here that the difficulty can be circumvented, at least for the Loschmidt echo.
 The idea is to perform the analytic continuation semi-numerically. We explain this procedure in the case the semi-infinite geometry $\ell=L_A \ll L_B,L$, and will
 use it extensively in the rest of this paper.

\begin{figure}[ht]
\begin{center}
\includegraphics{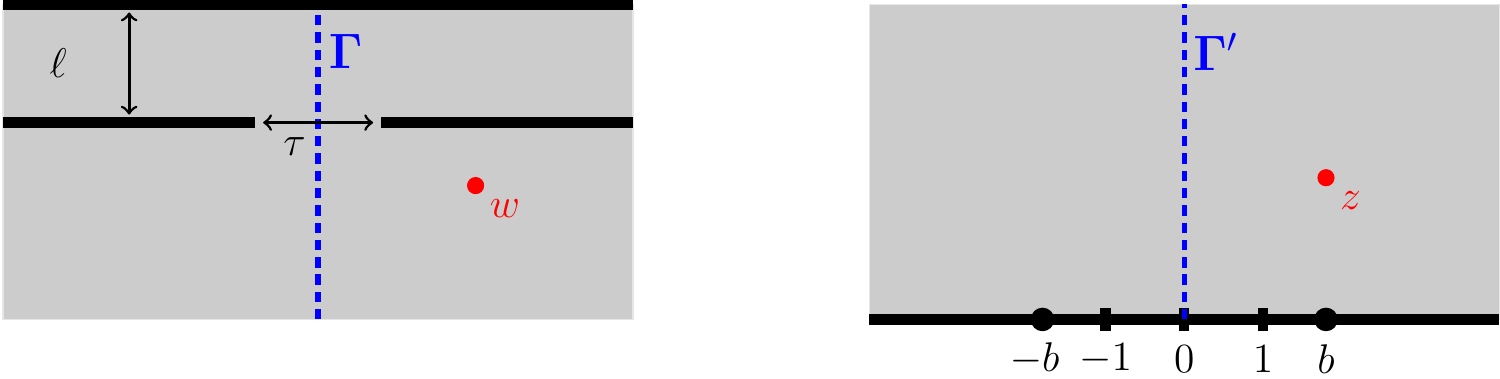}
\end{center}
\caption{Semi-infinite geometry (left) and upper half-plane $\mathbb{H}$ (right). The conformal transformation $w(z)$ is given by Eq.~\ref{eq:inf_pants}.}
\label{fig:inf_pants}
\end{figure}
We consider a subsystem of size $L_A = \ell$, coupled at time $t = 0$ to a semi-infinite chain. In imaginary time, we need the semi-infinite geometry shown in Fig.~\ref{fig:inf_pants}.
The Schwarz-Christoffel transformation mapping the upper half-plane to this semi-infinite geometry reads in this case
\begin{equation}\label{eq:inf_pants}
 w(z)=\frac{\ell}{\pi}\left[\frac{2b^2}{1-b^2}\frac{1}{z}+\log \left(\frac{z+1}{1-z}\right)\right]
\end{equation}
Our parametrization of $w(z)$ ensures that $w'(\pm b)=0$. The distance $\tau$ between the two slits is given by
\begin{equation}\label{eq:tau_inf}
 \tau(b)=w(b)-w(-b)=\frac{2\ell}{\pi} \left[\frac{2b}{1-b^2}+\log \left(\frac{b+1}{b-1}\right)\right],
\end{equation}
and the variation of the (logarithmic) Loschmidt echo $\mathcal{F} = - \log \mathcal{L}$ can be expressed as a residue of $f(z)=S(w,z)(dw/dz)^{-1}$ at $z=b$:
\begin{eqnarray}
 \delta \mathcal{F}&=&\frac{c}{6}\; {\rm Res}[f(z);z=b]\;\delta \tau\\
&=&\frac{c}{6}\; {\rm Res}[f(z);z=b] \;\frac{\delta \tau}{\delta b} \delta b
\end{eqnarray}
We get 
\begin{equation}\label{eq:freenrj_inf}
 \mathcal{F}(b) =\frac{c}{6}\left(\Re e \left[ \frac{2}{1-b^2} \right]+\frac{3}{2}\log |b| -\log |b^2-1|\right)
\end{equation}
where $\Re e [a]$ is the real part of $a$. This is important because we now want to go back in real time.
As already suggested, one important difference with the two preceding cases is that the equation (\ref{eq:tau_inf}) relating $\tau$ to $b$ cannot be explicitly inverted.
 The analytical continuation ($\tau \to i v_F t +\epsilon$) can nevertheless be performed numerically. For each time $t$, we solve
\begin{equation}\label{eq:it_inf}
 i v_F t+\epsilon=\tau(b)=\frac{2\ell}{\pi} \left[\frac{2b}{1-b^2}+\log \left(\frac{b+1}{b-1}\right)\right],
\end{equation}
 and the solution $b(t)$ can be injected in Eq.~\ref{eq:freenrj_inf}.
 Some further technicalities are presented in \ref{sec:trajectories}.
\subsection{Extension to arbitrary aspect ratios}
\subsubsection{Loschmidt echo}
\begin{figure}[ht]
\begin{center}
\includegraphics{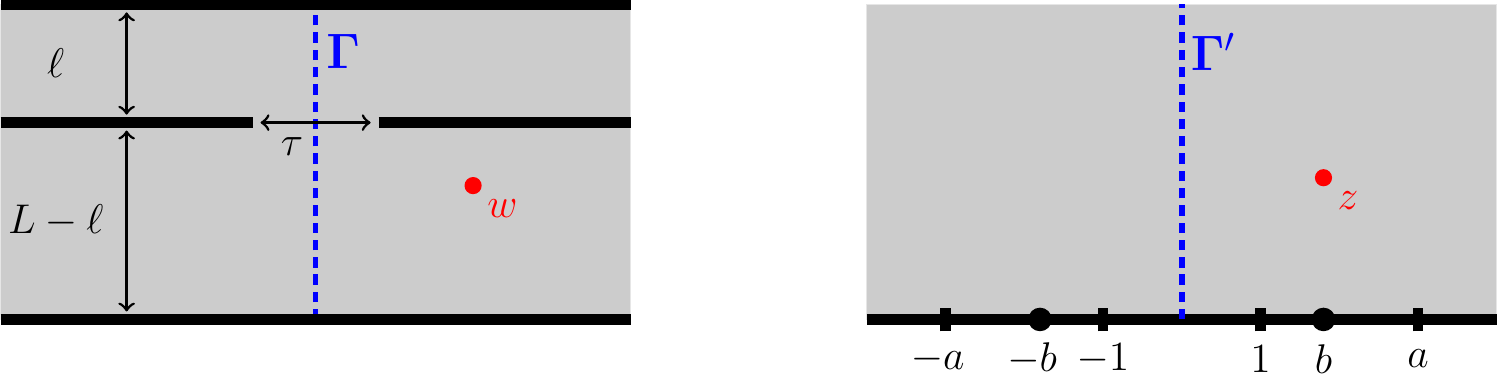}
\end{center}
\caption{``Asymmetric double-pants'' geometry (left) and upper half-plane $\mathbb{H}$ (right). The conformal transformation
$w(z)$ is given by Eq.~\ref{eq:conf_apants}.}
\label{fig:asymmetric_pants}
\end{figure}

The method developed in Sec.~\ref{sec:cft_ns_infinite} can be pushed further, to treat any aspect ratio  $x=\ell/L = L_A/(L_A+L_B)$. The conformal transformation which maps the
 upper half plane onto the ``asymmetric double-pants'' geometry in Fig.~\ref{fig:asymmetric_pants} is given by
\begin{equation}\label{eq:conf_apants}
 w(z)=\frac{L}{\pi}\left[x\log \left(\frac{z-1}{z+1}\right)+(1-x)\log \left(\frac{z-a}{z+a}\right)-i\pi\right]
\end{equation}
As before, we first need to determine the distance between the two slits $\tau=w(b)-w(-b)$, where $b$ is the solution of
$w'(\pm b)=0$. We find
\begin{eqnarray}\label{eq:tau_ns}
 b^2 &=& a\frac{xa+(1-x)}{(1-x)a+x}\\
\tau&=&\frac{2L}{\pi}\left[x \log \left(\frac{b-1}{b+1}\right) +(1-x)\log \left(\frac{a-b}{a+b}\right)\right]
\end{eqnarray}
Then the free energy can again be computed as a residue at $z=b$, and we get after some algebra
\begin{eqnarray}\label{eq:le_ns}
 \nonumber \mathcal{F}(a)&=&\frac{c}{6}\left[\frac{3}{4}\log |a| +\frac{3x^2-3x+1}{x(1-x)}\log |a+1|-\frac{x^2-x+1}{x(1-x)} \log |a-1| \right. \\ 
&& \left. \qquad +\, \frac{1}{4}\log |ax+(1-x)|+\frac{1}{4}\log |a(1-x)+x|\right]
\end{eqnarray}
When $x$ is the inverse of some integer ($x=1/p\;,p\in\mathbb{N}$), getting $a$ as a function of $\exp\left(\pi \tau/2L\right)$ amounts to
 finding the roots of a polynomial of degree $2p-2$. Doing so for $\tau=iv_F t+\epsilon$ we get a whole set of solutions, but only one of them corresponds to the actual real-time LLE.
\subsubsection{Entanglement entropy}
The entanglement entropy turns out to be much more complicated in the asymmetric geometries. As already stated, the calculation of the one-point function
 requires the inversion of the conformal mapping of Eq.~\ref{eq:conf_apants}. When $x=1/q$ where $q$ is some integer, inverting the mapping amounts
 to finding the roots of a polynomial of degree $q$. An analytic expression of those is in principle available only for $q=3$ and $q=4$. Such an analytic expression is needed in order
to perform the analytic continuation to real time. In practice, the expressions we get for the
 entanglement entropy are gruesome, and not particularly illuminating.
 We managed to treat the $x=1/3$ case, and present our result here for the sake of completeness. First we introduce the auxiliary function
\begin{equation}\label{eq:ee_ns_1s3_a}
\fl
 f_\epsilon(t)=\frac{\left\{
4(1-3\epsilon)\big[17-\cos (3\pi t)\big]\cos \left(\frac{3\pi t}{2}\right)-24i \sqrt{2}(1-\epsilon)\sqrt{1+16\epsilon-\cos (3\pi t)} \sin \left(\frac{3\pi t}{2}\right)
\right\}^{1/3}}{(4+2\epsilon)\sin\left(\frac{3\pi t}{2}\right)},
\end{equation}
We denote by $g_\epsilon=\Re {\rm e} (f_\epsilon)$ and $h_\epsilon=\Im {\rm m} (f_\epsilon)$
 the real and imaginary parts of $f_\epsilon$.
 We also set $v_FL=1$ for convenience. The EE can then be expressed using $f_\epsilon(t)$ and its derivative $f_\epsilon'(t)=g_\epsilon'(t)+ih_\epsilon'(t)$:
\begin{equation}\label{eq:ee_ns_1s3_b}
S(t)=\frac{c}{12}\log \left\{
\frac{ \left[h_\epsilon(t)\right]^2}{
\left[\frac{3\pi}{2} \csc^2 \left(\frac{3\pi t}{2}\right)+g_\epsilon'(t)\right]^2-
3\left[
h_\epsilon'(t)
\right]^2
}
\right\}+{\rm cst}.
\qquad ( \epsilon \ll t<2/3),
\end{equation}
where $\csc u=1/\sin u$. It is important to keep the regulator $\epsilon$ finite in Eq.~\ref{eq:ee_ns_1s3_b}, 
because the denominator is of order $\epsilon$. We also note that this expression can
 probably be further simplified. 
A comparison with numerical simulations is shown in Sec.~\ref{sec:plateaus}.
%%%%%%%%%%%%%%%%%%%%%%%%%%%%%%%%%%%%%%%%%%%%%%%%%
%%%%%%%%%%%%%%%%%%%%%%%%%%%%%%%%%%%%%%%%%%%%%%%%%
\section{Light-cone effects: plateaus of the LLE and non-universal decay of the entanglement entropy}
\label{sec:plateaus}
%%%%%%%%%%%%%%%%%%%%%%%%%%%%%%%%%%%%%%%%%%%%%%%%%
%%%%%%%%%%%%%%%%%%%%%%%%%%%%%%%%%%%%%%%%%%%%%%%%%

\begin{figure}[ht]
\begin{center}
\includegraphics{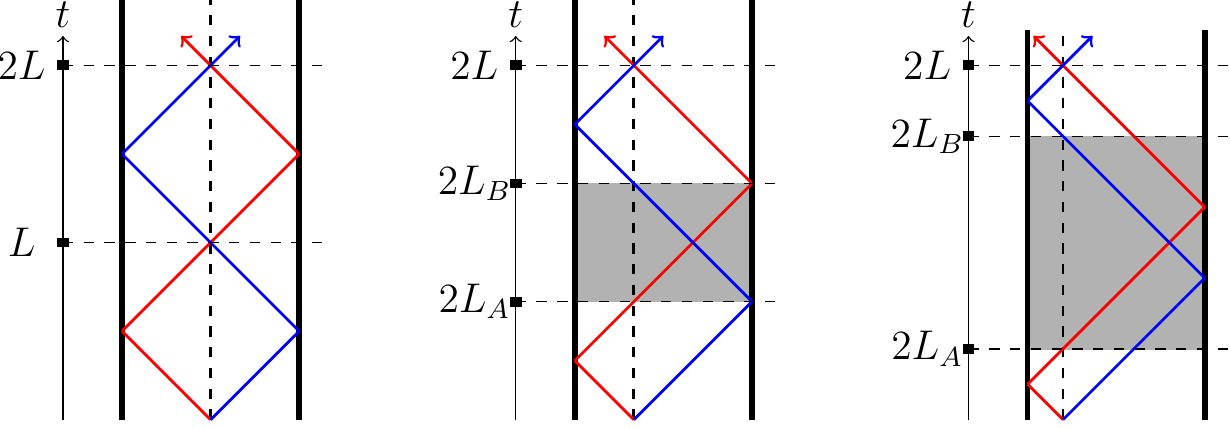}
\end{center}
\caption{Drawing of quasi-particles emitted at $t=0$ by the local quench. They propagate at the speed of light $v_F$
 and bounce against the wall. Left : $x=1/2$, centre : $x=1/3$, right : $x=1/5$. The plateau regions (filled in grey) correspond
 to the case when the two quasiparticles are in the same subsystem.}
\label{fig:particles2}
\end{figure}
We wish now to interpret some of the results we derived for the Loschmidt echo, as well as for the entanglement entropy.
 As emphasized in the introduction, we are focusing here mainly on finite-size effects. This is especially interesting in the
case of a local quench, because the quasiparticles emitted at $t=0$ can bounce back on the boundaries. Now something interesting
 happens when the left-right periodicity is broken $L_A\neq L_B$: for certain periods of time, all the quasiparticles can be in
 the same subsystem (see Fig.~\ref{fig:particles2}). Since they are supposed to transport information about entanglement and correlations, it is reasonable
 to expect that the behaviour of both the (logarithmic) Loschmidt echo and the EE will drastically change. 
This is indeed the case, both quantities exhibit light-cone effects. However, we will see that they are quite different.
 For the EE we see a slow non symmetric decay, whereas we get striking plateaus for the LLE.

\paragraph{}

We also wish to compare our analytic results to several numerical simulations in free fermionic spin chains, which 
are described in the continuum limit by CFT with central charges $c=1$ or $c=1/2$. This section is organized as follows. First
 we give some details about the model -- so called XY-chain in transverse field -- and show how the quasiparticles can
 be identified. We then discuss separately the behaviour of the entanglement entropy and the Loschmidt echo. Finally we
 introduce a variant of the Loschmidt echo (``detector geometry'') specifically constructed to probe the light-cone effect, and discuss the results.
\subsection{XY spin chain in transverse field}
We consider the following Hamiltonian
\begin{equation}
 H=-\frac{1}{2}\sum_{j=1}^{L-1}\left[\left(\frac{1+r}{2}\right) \sigma_j^x \sigma_{j+1}^x +\left(\frac{1-r}{2}\right)\sigma_j^y \sigma_{j+1}^y\right]
-\frac{h}{2} \sum_{j=1}^L \sigma_j^z, 
\end{equation}
where the boundary conditions for the spins are free and $L$ is assumed to be even.
 As is well-known, $H$ can also be expressed as a quadratic form in fermionic
 operators, which makes it one of the simplest exactly solvable spin chains. The fermionic structure also allows to
 calculate numerically both the entanglement entropy and the Loschmidt echo for very large system sizes.
 The phase diagram is quite rich. In particular there are two critical lines, lying in different universality classes. 
 
\subsubsection{$XX$ chain: free boson critical line}

\label{sec:filling}

This corresponds to $r=0$ and $-1<h<1$. $H$ can in this case be rewritten up to
 a unimportant constant as
\begin{equation}
 H=\sum_{m=1}^{L} \epsilon_m c_m^\dag c_m\quad,\quad \epsilon_m=-h-\cos \left(\frac{m\pi}{L+1}\right),
\end{equation}
 where we have the usual anticommutation relations $\{c_m,c_n^\dag\}=\delta_{mn}$ and $\{c_m,c_n\}=\{c_m^\dag,c_n^\dag\}=0$ for the fermions.
 The quasi-momenta are the $k_m=m\pi/(L+1)$ and the Brillouin zone extends over the interval $(0,\pi)$. At zero magnetic field $h=0$, the ground-state
 can simply be obtained by filling half of the first quasi-momenta:
\begin{equation}
 E_0(L)=-\sum_{m=1}^{L/2}\cos \left(\frac{m\pi}{L+1}\right)
\end{equation}
The asymptotics of $E_0$ when $L\to \infty$ can be accessed using the Euler-Maclaurin formula:
\begin{equation}
 E_0(L)=\alpha L +\beta -\frac{\pi}{24 L}+\mathcal{O}(1/L^2)
\end{equation}
$\alpha$ and $\beta$ are non universal constants and direct comparison with Eq.~\ref{eq:finitesizec}, yields $v_F \times c=1$. The Fermi
 velocity can simply be calculated from the dispersion relation, $v_F=|\frac{d \epsilon}{dk}|_{|_{k=k_F=\pi/2}}=1$ and we get the textbook result that
 the XX chain lies in the $c=1$ (free boson) universality class. As already explained, we expect that the local quench will mainly generate low-energy
 excitations, therefore at energies close to the Fermi level. In this region, the dispersion relation can be linearised, and this is precisely
 what the CFT describes. 
 
\paragraph{}
At non-zero magnetic field, the filling fraction is not $1/2$ anymore and the number of fermions is an integer $n\neq L/2$.
 This case turns out to be trickier. 
 Let us focus for the moment on the case when $n=\rho L$, where $\rho$ is some rational number. The ground-state energy is
\begin{equation}
 E_0(L)=-\sum_{m=1}^{\rho L}\cos\left(\frac{m \pi}{L+1}\right),
\end{equation}
 and applying the Euler-Maclaurin formula we get ($v_F=\sin k_F\sim \sin \rho \pi$)
\begin{equation}
 E_0(L)=\alpha' L+\beta'-\frac{\pi v_F}{24 L}\left[1-12(\rho-1/2)^2\right]+\mathcal{O}(1/L^2)
\end{equation}
To understand this result and its implications, we follow Zagoskin and Affleck \cite{Fermiedge}. Let us first define our convention
 for the Fermi level. We choose to place it between the last occupied and the first empty wave-vector
\begin{equation}
 k_F=\pi \frac{n+1/2}{L+1}=\pi \frac{\rho L+1/2}{L+1}
\end{equation}
Except at half-filling $\rho=1/2$, the particle-hole symmetry is not exact on the lattice: $\epsilon(k_n)-\epsilon(k_F)\neq \epsilon(k_F)-\epsilon(k_{n+1})$.
 This effect has surprisingly important consequences in the continuum limit. In a bosonisation picture, it acts as a phase shift\cite{Fermiedge}
 and changes the effective boundary conditions on the boson field. 
\paragraph{}

As explained in Ref.~\cite{Fermiedge}, this difficulty can usually be circumvented by choosing
 appropriate conventions for the length of the chain and the number $n$ of fermions. For example at quarter-filling, choosing the length of the chain as
$L=4n+1$ restores the particle-hole symmetry and therefore cancels the anomalous term $12 \times (1/4-1/2)^2$.
This yields the desirable result $c=1$ for the central charge. In this case the boundary conditions at both ends
 of the chain are the same, which is easier to treat from a CFT perspective. However, we are interested here in the effect of local quench,
 where the interaction between two subsystems $A$ and $B$ is switched on. For parity reasons this trick cannot be applied to the chains $A$,
 $B$ and $A\cup B$ at the same time. In other words, it is not possible (away from half-filling) to maintain the same boundary condition everywhere
 and we must introduce boundary changing operators in the CFT calculation. It is known that they don't affect the leading term in the entanglement entropy,
 and we therefore expect our formulas to hold away from half-filling in that case. However, these operators can have a noticeable effect on the
 scaling behaviour of the (logarithmic) Loschmidt echo and we expect our results to break down away from half-filling. Although these effects can be treated by
 standard CFT techniques, we do not study them in this paper. Therefore, we will stick to numerical checks at half-filling. Finally, let us mention
 that this discussion also applies to the XXZ chain in a magnetic field. There is also a phase shift away from half-filling. It takes a more complicated form
which has been computed using Bethe-ansatz techniques\cite{Heisenberg1,Heisenberg2}.
\subsubsection{Ising critical line}
\label{sec:Ising}
\begin{figure}[ht]
\includegraphics[width=8cm]{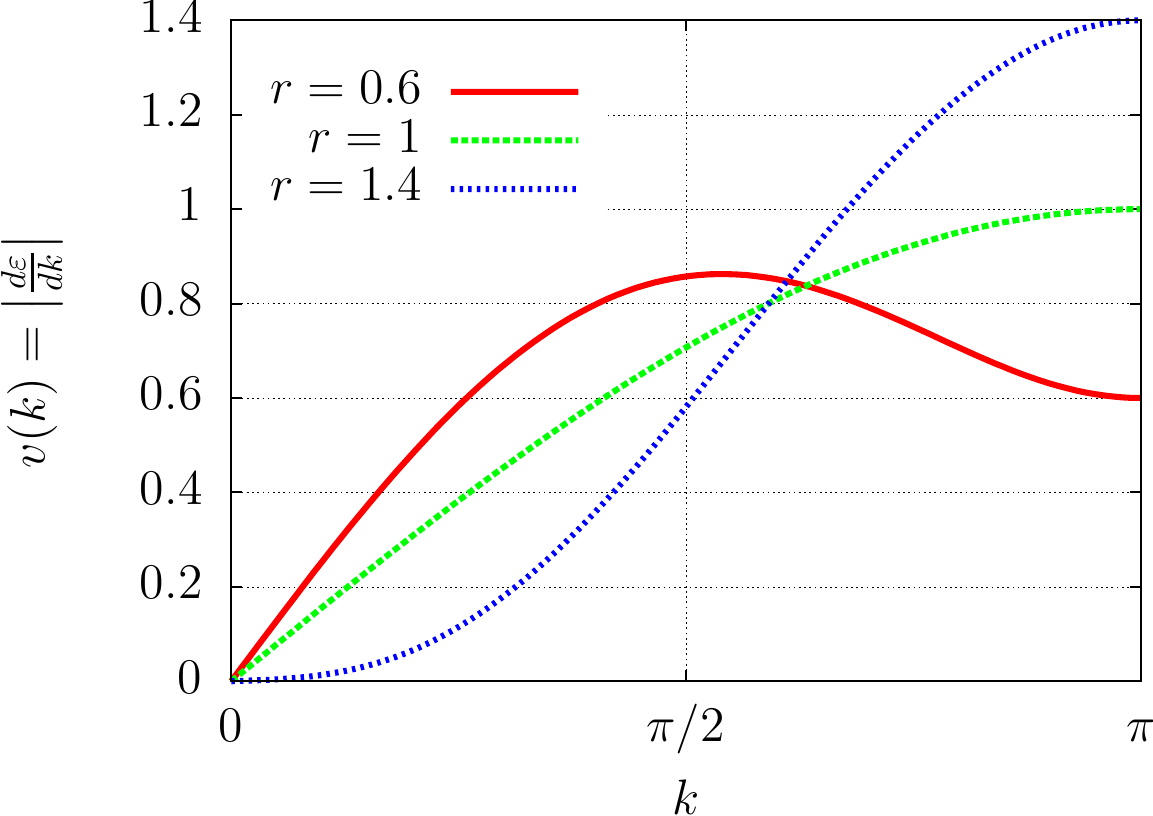}
\includegraphics[width=8cm]{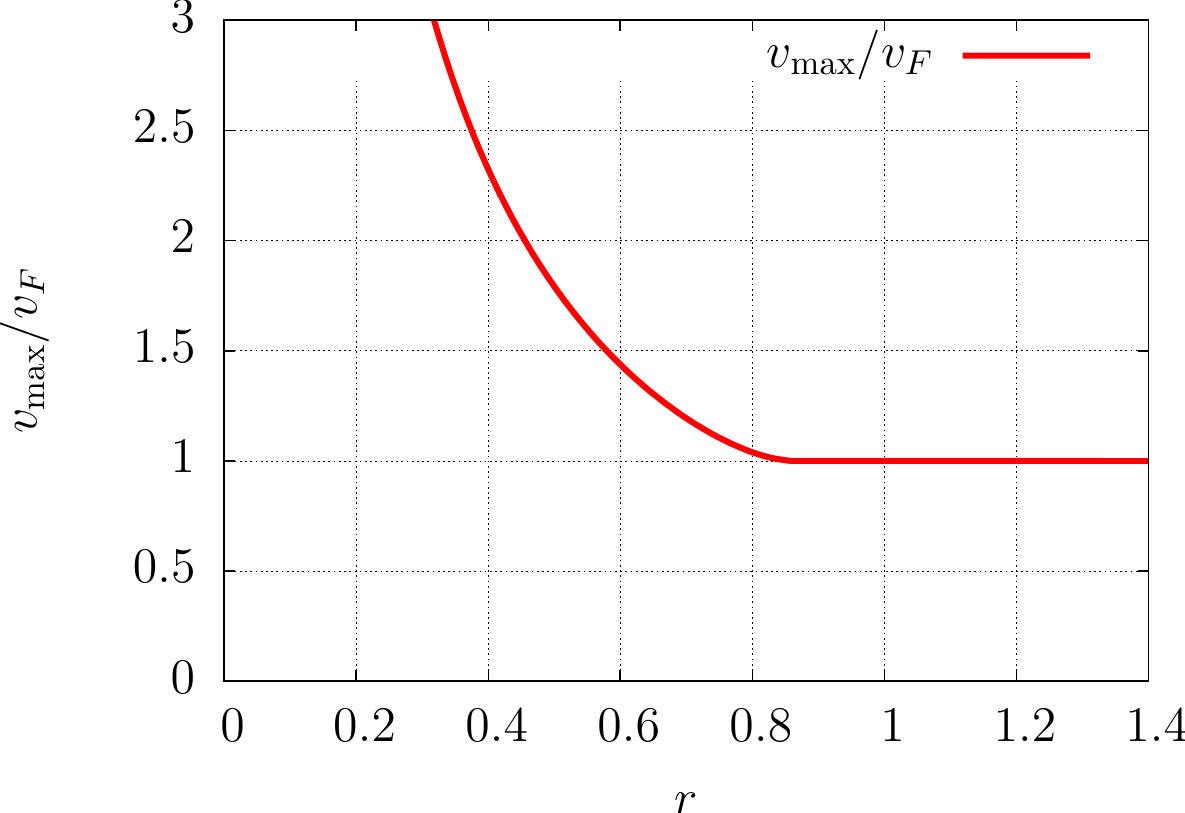}
\caption{Left panel: group velocity $v(k)=|d\varepsilon/dk|$ for three different values of $r$ ($r=0.6,1,1.4$). $v_{\rm max}$ is the maximum group velocity allowed.
 Right panel: ratio $v_{\rm max}/v_{F}$, where $v_F$ is the Fermi velocity $v_F=v(k=\pi)$. For $r<\sqrt{3}/2$, the ratio is greater than one.}
\label{fig:dispersion}
\end{figure}

 This corresponds to $h=1$ and $r>0$. In this case the Hamiltonian can be diagonalized by a Bogoliubov transformation:
\begin{equation}
 H=\sum_{m=1}^{L} \varepsilon_m \left(d_m^\dag d_m -1/2\right)
\end{equation}
and the one-particle energies are given by
\begin{equation}
 \varepsilon_m=\sqrt{\left(h+\cos k_m\right)^2+r^2 \sin^2 k_m}
\end{equation}
The wave-vectors $k_m$ can be obtained as solutions of an implicit equation, and lie in the interval $(0,\pi)$. For $r=1$ we have the Ising chain in transverse
 field, and they take a simple form:
\begin{equation}
 k_m=\frac{m \pi}{L+1/2}\quad, m=1,\ldots,L.
\end{equation}
This model is known to have central charge $c=1/2$ for all values of $r>0$. In the continuum limit, the dispersion relation is
\begin{equation}
 \varepsilon(k)=\sqrt{(1+\cos k)^2+r^2 \sin^2 k},
\end{equation}
the Fermi momentum is at $k_F=\pi$, the Fermi velocity is $v_F=d\varepsilon/dk|_{k=\pi}=r$, and the CFT again describes
 excitations close to the Fermi point. For $r<\sqrt{3}/2$, something interesting happens: $\varepsilon(k)$ is not convex
 and the Fermi speed $v_F$ is not the fastest speed available for the excitations (see Fig.~\ref{fig:dispersion} for an example).
This maximum speed
\begin{equation}
 v_{\rm max}=\underset{k}{{\rm max}} \;v(k)=\underset{k}{{\rm max}} \;\left|\frac{d\varepsilon}{dk}\right|
\end{equation}
 is sometimes called the Lieb-Robinson speed\cite{LiebRobinson} and corresponds to the maximum velocity allowed for the
propagation of a signal in the quantum system. Its role in the context of global quenches has been emphasized\cite{Revivals,mathey2010,eegcxy}.
As we shall see, these fast particles ---which propagate faster than the 
CFT particles discussed in the introduction--- affect both the entanglement entropy and the Loschmidt echo.
 Conversely, in the case $r>\sqrt{3}/2$, the Fermi speed is the fastest available speed,
 and the behaviour of both the EE and the Loschmidt echo is rather affected by particles which
propagate slowly (compared to $v_F$).

\subsection{Non-universal decay of the entanglement entropy}
\begin{figure}[ht]
 \begin{center}
\includegraphics[width=7.7cm]{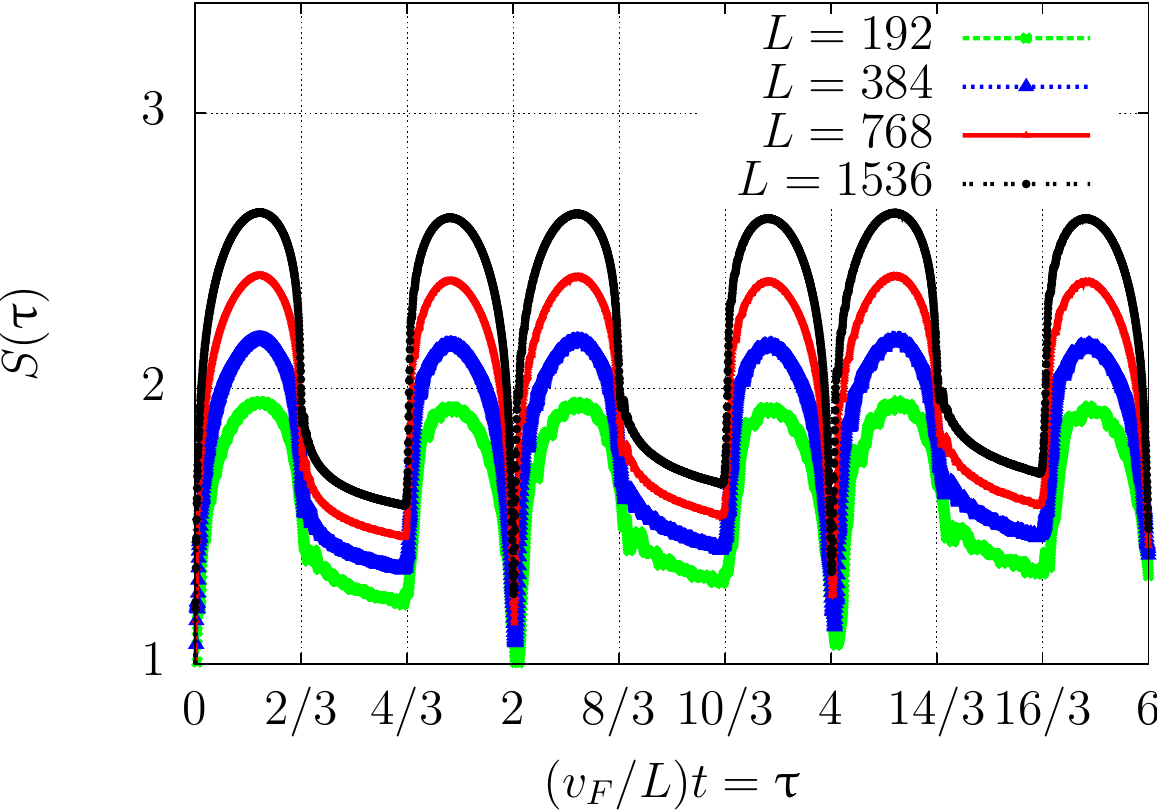}
\includegraphics[width=7.7cm]{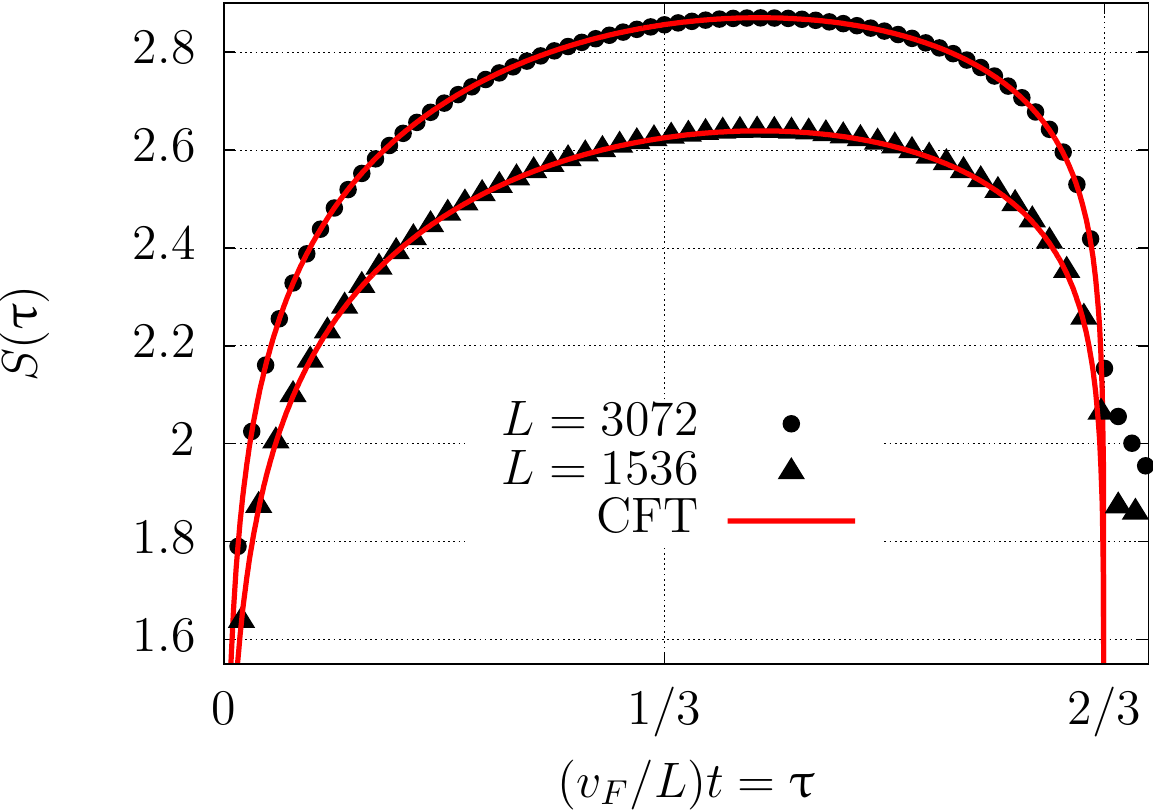}
 \end{center}
\caption{Entanglement entropy for the aspect ratio $x=1/3$. Left panel : Numerical simulations in the XX chain, shown against the rescaled time $(v_F/L)t$.
 Right panel : zoom on the ``CFT region'' $0< t <2L/(3v_F)$, and comparison with the analytical calculation of Eqs.~\ref{eq:ee_ns_1s3_a} and \ref{eq:ee_ns_1s3_b}.}
\label{fig:ee_1s3}
\end{figure}

\begin{figure}[ht]
 \includegraphics[width=8cm]{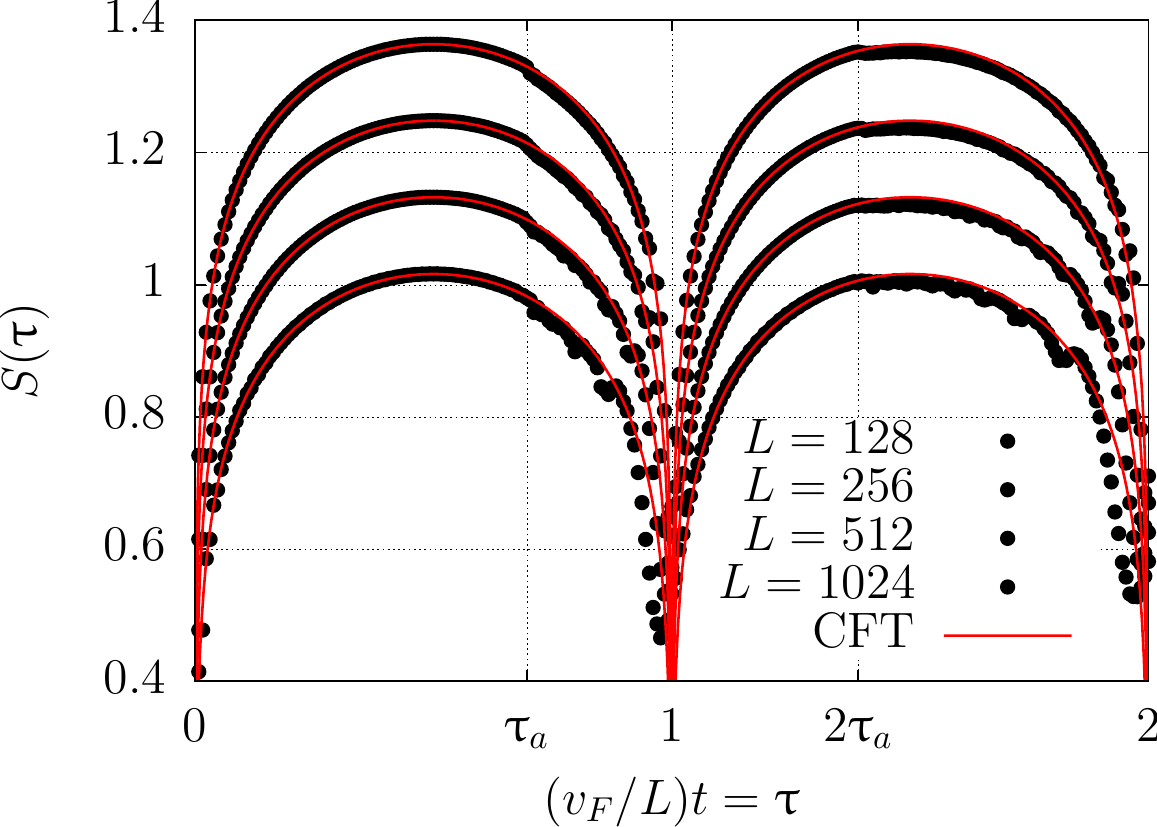}
\includegraphics[width=8cm]{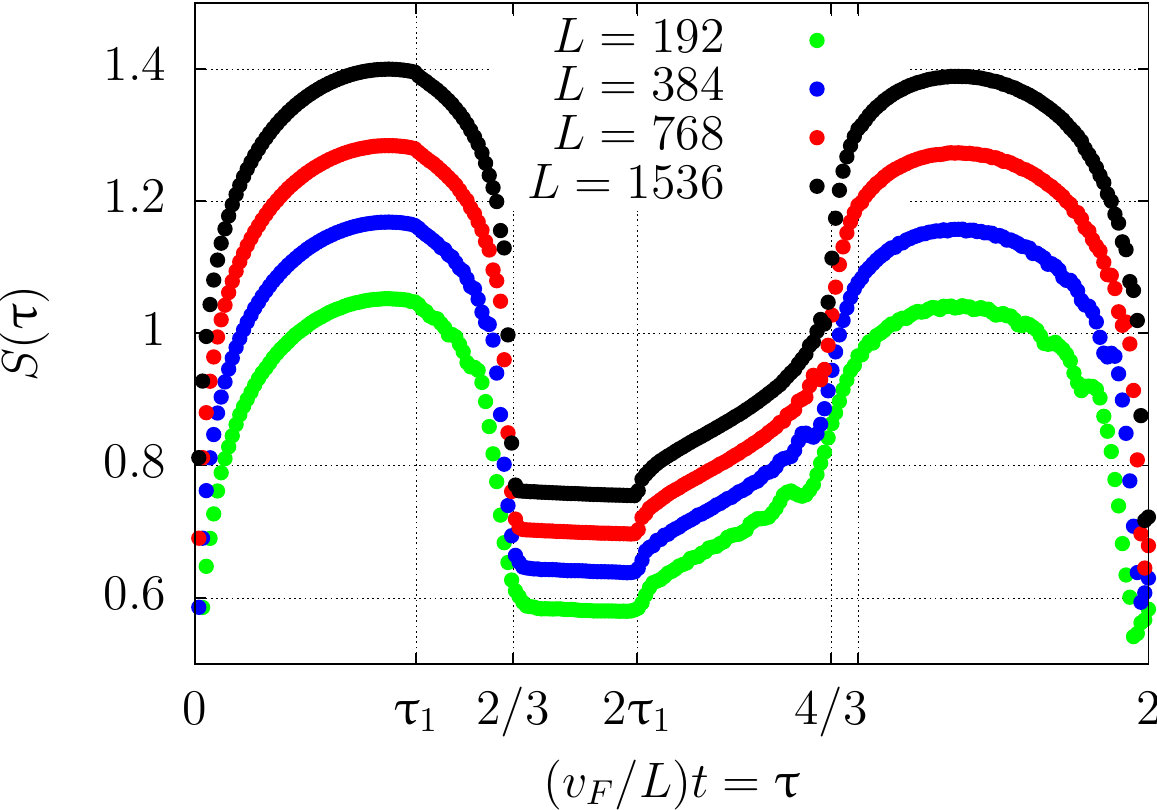}
\caption{Left panel: Entanglement entropy as a function of 
the rescaled time $\uptau=(v_F/L)t$ for the ratio $x=1/2$, the anisotropy $r=0.6$ and for systems sizes $L=128,256,512,1024$.
 At rescaled times $\uptau_a=v_F/v_{\rm max}$ and $2\uptau_a$ the CFT calculation breaks down, although it remains a good approximation. Right panel:
 Entanglement entropy for $x=1/3$, $r=0.6$ and for system sizes $L=192,384,768,1536$. The rescaled times of interest are $\uptau_1=2 v_F/(3 v_{\rm max})$, $2\uptau_1$ and $3\uptau_1$. }
\label{fig:eer}
\end{figure}

\paragraph{}
We first focus our attention on the EE. The only asymmetric finite-size system we have been
able to treat corresponds to
the aspect ratio $x=L_A/L=\ell/L=1/3$. The solution given by Eqs.~\ref{eq:ee_ns_1s3_a} and \ref{eq:ee_ns_1s3_b}
 can be compared to numerical simulations in free fermionic systems. 
 Fig.~\ref{fig:ee_1s3} shows such a comparison
 in the XX chain for very large system sizes, up to $L=3072$. 
The agreement between the two is excellent at the beginning,
 but suddenly breaks down precisely at the time ($t=2L_A/v_F$) when the two light-cone boundaries (or position of left-moving and right-moving CFT quasi-particles) lie both in subsystem B (see Fig.~\ref{fig:particles2}
 in the middle). Then the entropy slowly decays up to time $t=2L_B/v_F$, when the CFT calculation becomes accurate again. 
 Such a decay was already observed in the semi-infinite geometry in Ref.~\cite{EislerPeschel2008}, also for the XX chain.
 It can be attributed to the slower quasi-particles progressively entering the subsystem $B$. 
\paragraph{}
 Several interesting observations can be made:
\begin{itemize}
 \item Global scaling with the system size $L$: in the ``CFT region'' (corresponding to $t< 2L_A/v_F$ and $t> 2 L_B/v_F$ in the first period) the numerical simulations show a global
 scaling $(c/3) \log L$. This is a consequence of the presence of two slits in the CFT setup, which contribute both\cite{CalabreseCardy2004}
 to a $(c/6) \log L$ for open chains. This is not true in the ``decay region'' ($2L_A/v_F<t<2L_B/v_F$) where the global scaling is half of the previous one: $(c/6) \log L$. This can be interpreted as
 follows. From the point of view of the subsystem $A$, there are no CFT quasi-particles anymore in $A$ and thus $A$ sees ---up to non-universal contributions---
 a wave-function close to the ground-state wave function $|A\cup B\rangle$ of the total system, which contributes exactly to a $(c/6) \log L$.
\item At $t=2 L_B$, the CFT particles are again in their own subsystem and contribute to the EE as predicted by CFT. However, one could expect 
non universal contributions to deteriorate the agreement with the numerics. This effect seems to be
negligible (Fig.~\ref{fig:ee_1s3}).
\end{itemize}
To test the universality of the EE, and the effect of the non-CFT quasiparticles, we also performed numerical checks in the XY chain
 in transverse field, for various values of the $r$ parameter in the Ising critical line. As explained in the previous section, the case
 $r<\sqrt{3}/2$ is particularly interesting, and puts the CFT calculation to a more serious test. Some results are presented in Fig.~\ref{fig:eer}, for
 $r=0.6$ and the aspects ratios $x=1/2$ and $x=1/3$.
 \paragraph{}
As explained in Sec.~\ref{sec:Ising}, the maximum speed is not the Fermi speed in this case, and some quasi particles will therefore
 enter the subsystem $B$ before  their CFT counterpart do. Although the effect of these fast particles should be small, it does not seem to disappear in the thermodynamic limit. 
 For example in the symmetric case $x=1/2$, there is a priori no decay region, and one expects the EE to be periodic with period $T=L/v_F$. However, at time $t=L/v_{\rm max}$
 the fast particles suddenly change subsystem, and a small deflagration ---analogous to the one observed in \cite{eegcxy} for a global quench--- can clearly be seen on the EE. The CFT calculation does not seem to hold anymore, due to these
 non universal effects. The same phenomenon happens at time $t=2L/v_{\rm max}$. For the aspect ratio
 $x=1/3$ the same interpretation holds. Their influence is however quite visible in the decay region.

\subsection{Plateaus of the LLE}
\begin{figure}[!ht]
 \begin{center}
  \includegraphics[width=8cm]{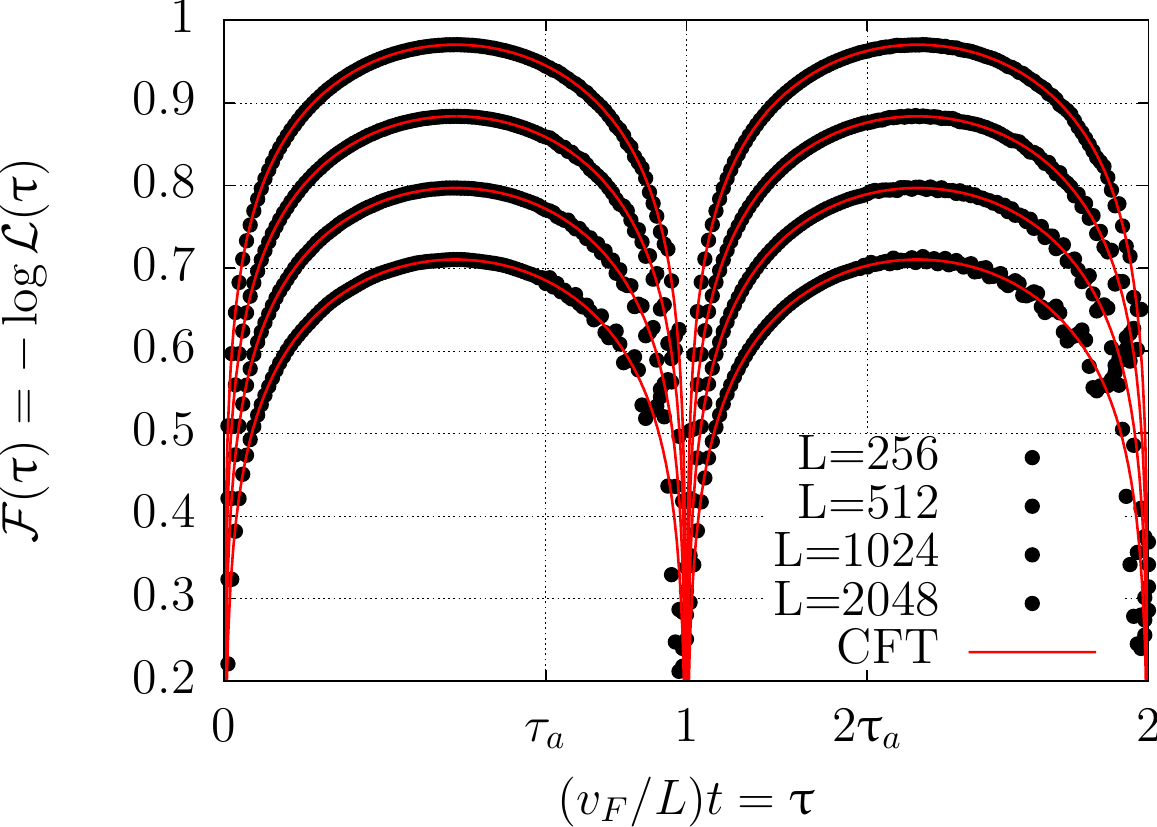}
  \includegraphics[width=8cm]{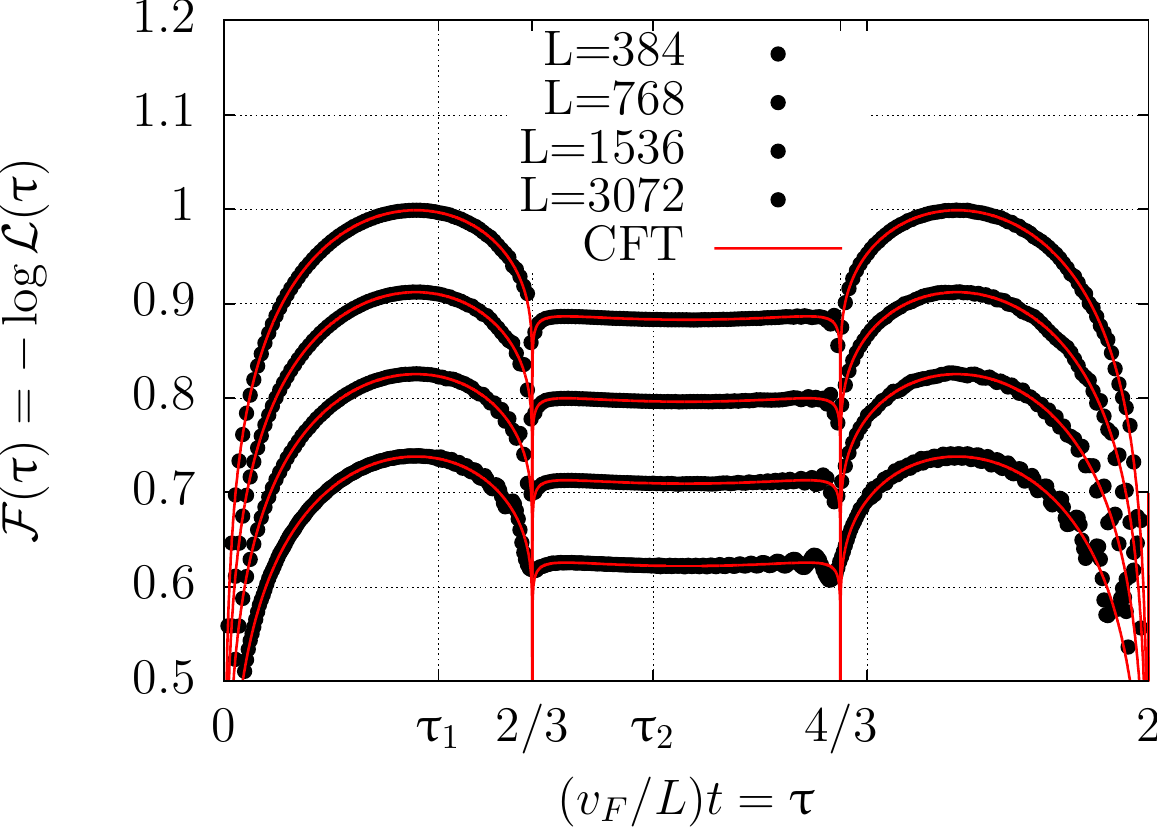}
 \end{center}
\caption{Left panel: Logarithmic Loschmidt echo as a function of rescaled time $\uptau=(v_F/L)t$ for the aspect ratio $x=1/2$, the anisotropy $r=0.6$ and
 for system sizes $L=256,512,1024,2048$. Right panel:
 Logarithmic Loschmidt echo as a function of rescaled time $\uptau=(v_F/L)t$ for the aspect ratio $x=1/3$, the anisotropy $r=0.6$ and
 for system sizes $L=384,768,1536,3072$. In both cases, finite size effects decrease for bigger and bigger system sizes. Note that,
 contrary to the entanglement entropy, the presence of fast particles ($v_{\rm max}>v_F$) does not spoil the plateaus.}
\label{fig:ler}
\end{figure}

\begin{figure}[ht]
\begin{center}
 \includegraphics[width=8.2cm]{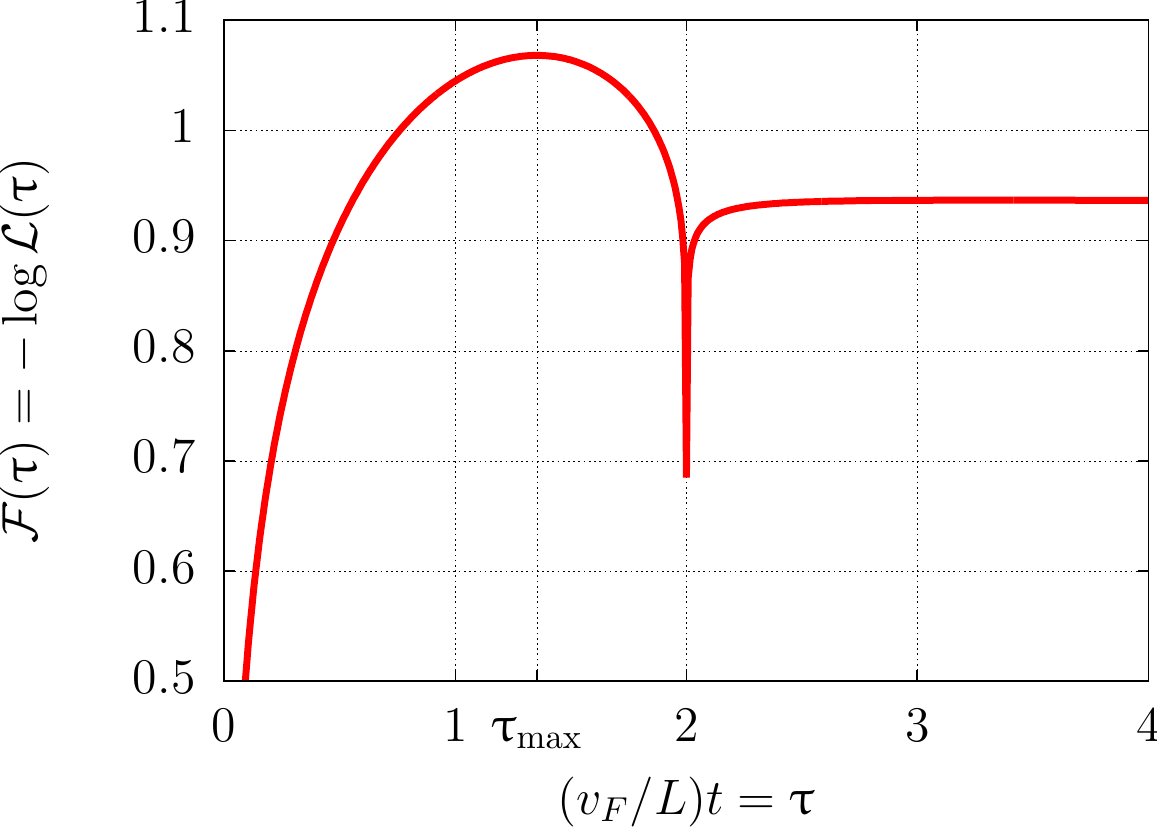}
 \includegraphics[width=8.2cm]{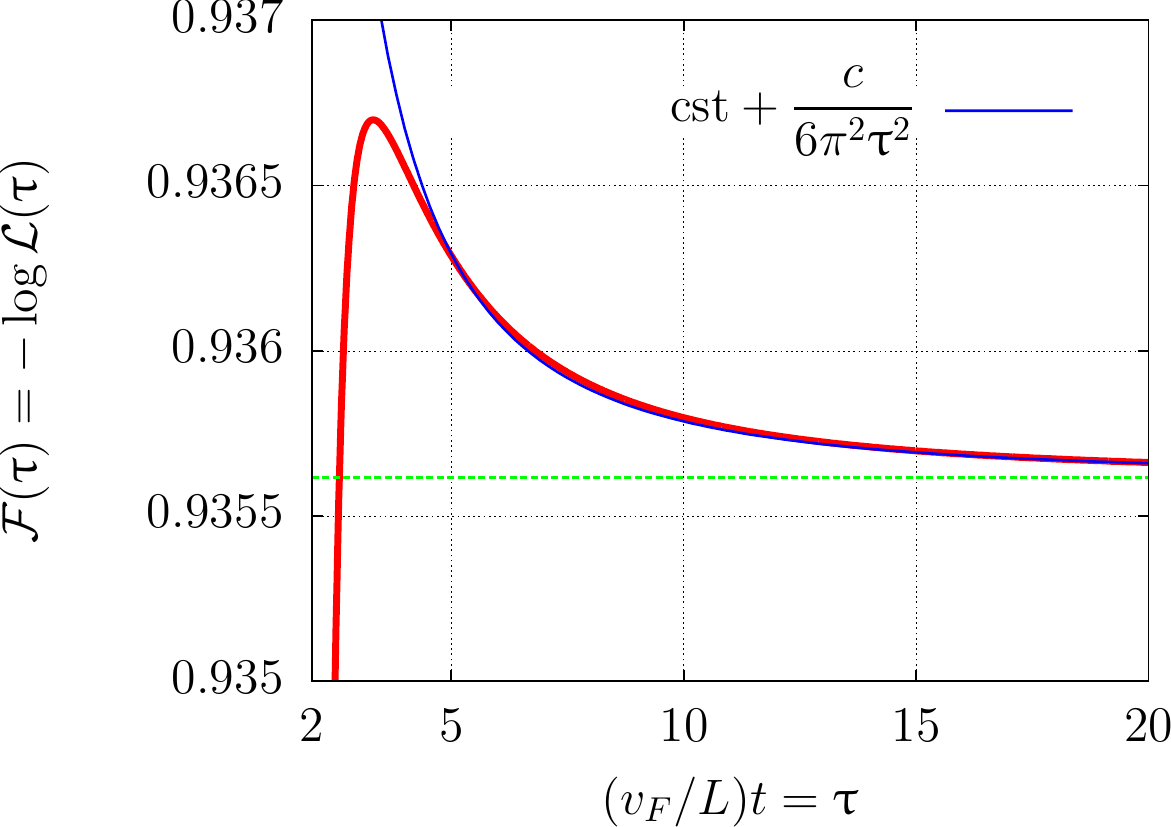}
\end{center}
\caption{Left panel: (logarithmic) Loschmidt echo $\mathcal{F}(t) =-\log \mathcal{L}(t)$ for the semi-infinite geometry and central charge $c=1$. Right panel: zoom on
 the plateau, for $v_F t/L>2$. The red curve is $\mathcal{F}(t)$ and the thin blue curve is the asymptotic expansion of Eq.~\ref{eq:le_inf_asymptotics}.}
\label{fig:loschmidt_inf}
\end{figure}

We now turn to the (logarithmic) Loschmidt echo. In the symmetric case $L_A = L_B$, its behaviour is very similar to the EE, and the two quantities are given by a very similar formula (\ref{eq:le_sym})-(\ref{eq:s_sym}).
For an asymmetric system $L_A \neq L_B$ such a close similarity between the EE and the LLE does not hold anymore, especially when
 the two quasiparticles are in the same subsystem (\textit{i.e.} the ``decay region''). The behaviour of the LLE is particularly striking: we get very flat plateaus, which
 can be accurately calculated in CFT (Eqs.~\ref{eq:tau_ns} and \ref{eq:le_ns}). 
\begin{figure}[!ht]
\begin{center}
 \includegraphics[width=8cm]{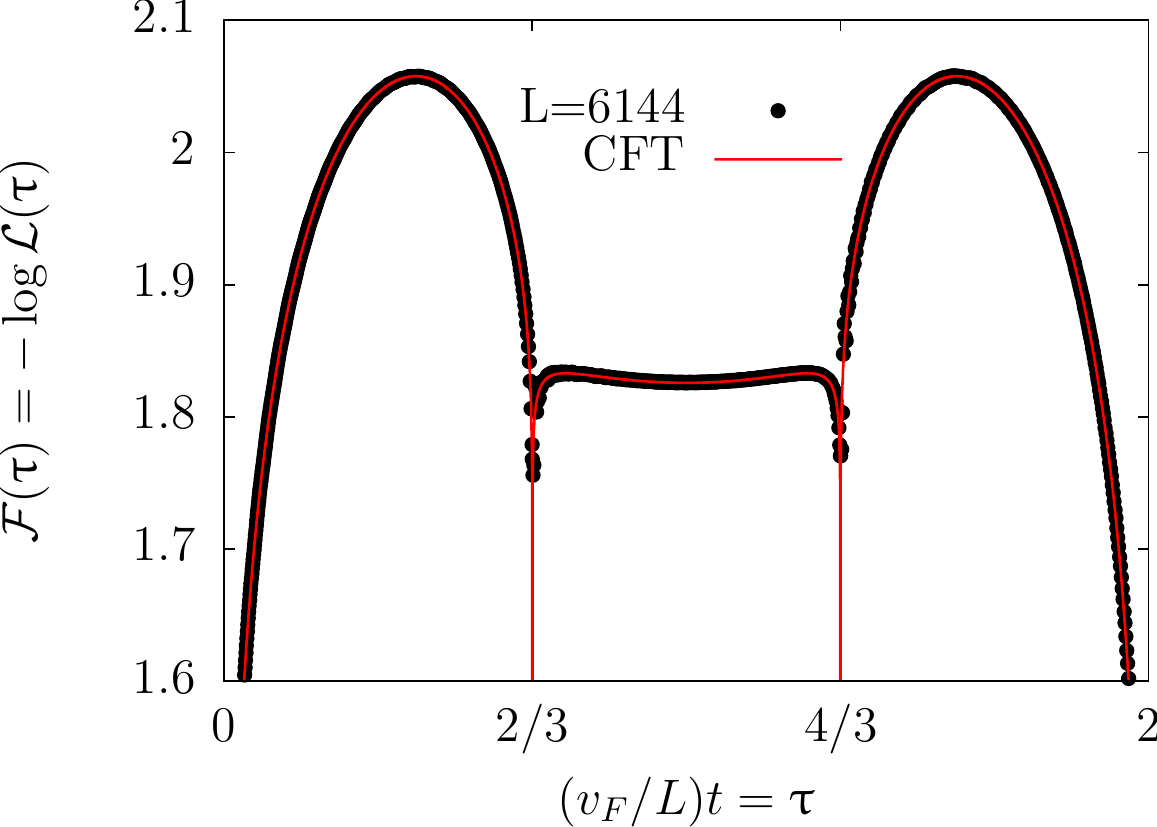}
 \includegraphics[width=8cm]{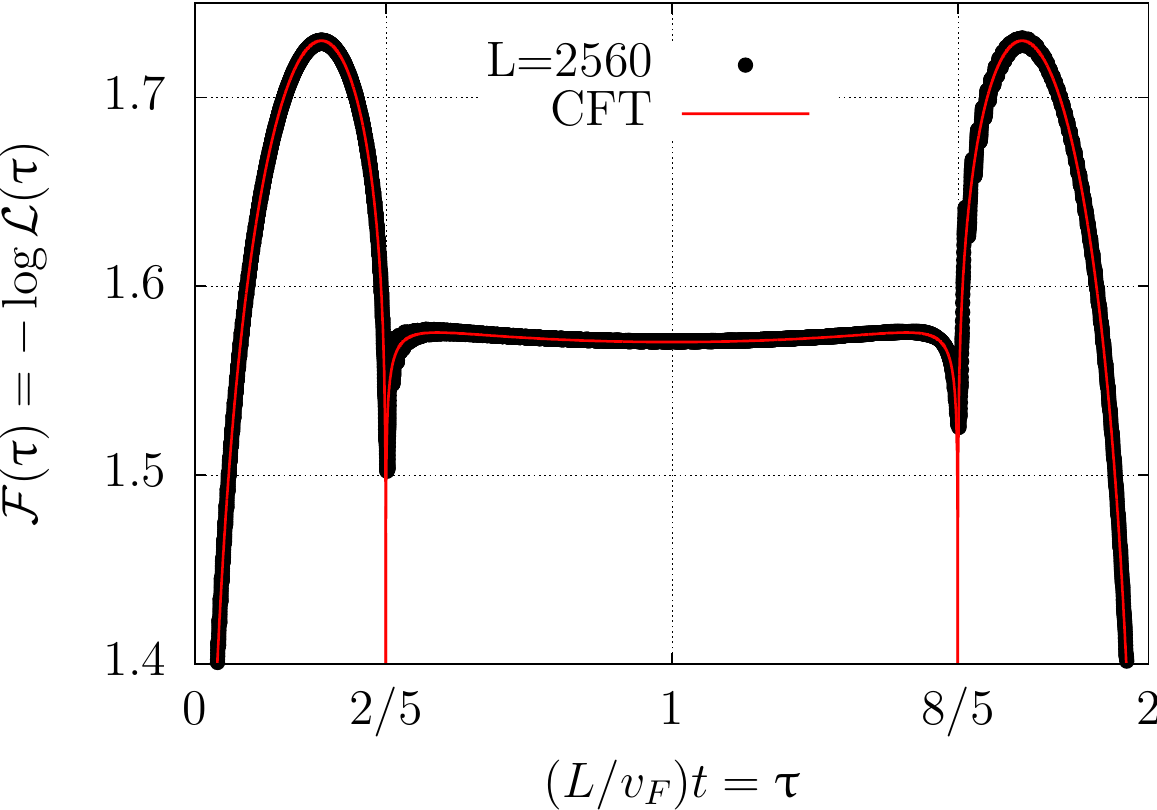}
\end{center}
\caption{Logarithmic Loschmidt echo $\mathcal{F}(t) = - \log \mathcal{L}(t)$ for the aspect ratios $x=L_A/(L_A+L_B)=1/3$ (left panel) and $x=1/5$ (right panel).
 Red curve : CFT calculation. Black points : numerical calculation in the XX chain.}
\label{fig:loschmidt_1s5}
\end{figure}
 See for example Fig.~\ref{fig:loschmidt_1s5}, for a comparison with the numerics in the cases $x=\ell/L=L_A/(L_A+L_B)=1/3$ and $x=1/5$.
Contrary to the EE, these plateaus are also symmetric with respect to $t\longmapsto 2L/v_F - t$, and have the same global scaling $(c/4) \log L$ as the other parts
 of the curve. To check the universality of our calculation, we also performed numerical checks in the XY chain in the case $v_{\rm max}>v_F$. The results are shown in
 Figure~\ref{fig:ler} for the aspect ratios $x=1/2$ and $x=1/3$. 
Interestingly, the Lieb-Robinson quasiparticles have a different effect on the LLE, compared to the EE. Finite size effects are enhanced when they change subsystem, but
 these diminish when increasing $L$. For very big system sizes the numerical calculations match very well our analytical prediction, even in the ``plateau region''. It therefore seems
 that the Loschmidt echo is less sensitive to non-universal effects than the EE. 
\paragraph{}
It is also interesting to study the semi-infinite geometry $L_A=\ell \ll L = L_A+L_B$. In this case the period is infinite, and one gets a single plateau
 starting at time $t=2\ell/v_F$. Although the analytical solution is given in implicit form, we can extract some interesting quantitative features about
 the small time and long time behaviour of the LLE. $\mathcal{F}(t) = -\log \mathcal{L}(t)$ is shown in Fig.~\ref{fig:loschmidt_inf}. 
We discuss now this curve (Eqs.~\ref{eq:freenrj_inf} and \ref{eq:it_inf}) in some detail. At small time $t\ll \ell$, we recover the expected behaviour \cite{DS}
\begin{equation}
 \mathcal{F}(t)=\frac{c}{4}\log t + {\rm cst}.
\end{equation}
This is simply a consequence of the fact that the quasiparticles do not feel the left boundary yet, and behave as if the system was infinite. 
Then $\mathcal{F}(t)$ keeps growing, up to a certain time when it reaches a maximum. This time $t_{\rm max}$ can even be calculated exactly
\begin{equation}
 t_{\rm max}=\frac{2}{\pi}\left[ \frac{\sqrt{2\sqrt{3}-3}}{\sqrt{3}-1}+\arctan \left(\frac{\sqrt{2\sqrt{3}-3}}{2-\sqrt{3}}\right)\right]\frac{\ell}{v_F}
\simeq 1.3538 \times\frac{\ell}{v_F}
\end{equation}
Also, the long time behaviour can be extracted
\footnote{For $t\to \infty$, the solution $a(t)$ of Eq.~\ref{eq:it_inf} can be written as $a(t)=\frac{b(t)+1}{b(t)-1}$, where
 $b(t)=\alpha(t)-i\beta(t)$ is a complex number with very small modulus. One can then obtain the following asymptotic expansions:
 $\beta(t)\sim \frac{1}{\pi(v_F t/\ell+1)}$ and $\alpha(t)\sim-2 \beta(t)^2 \log \beta(t)$.
 Injecting this in Eq.~\ref{eq:freenrj_inf} and expanding around $a=0$, we get the result.}
\begin{equation}\label{eq:le_inf_asymptotics}
  \mathcal{F}(t)- \mathcal{F}(\infty) = \frac{c}{6\pi^2 (v_F t/\ell)^2}+\mathcal{O}((v_F t/\ell)^{-3})
\end{equation}
The decay is algebraic, with a small ---but universal--- prefactor. This is actually the main reason why the plateau is so flat.

\subsection{``Detector'' geometries}
\begin{figure}[ht]
\begin{center}
\includegraphics{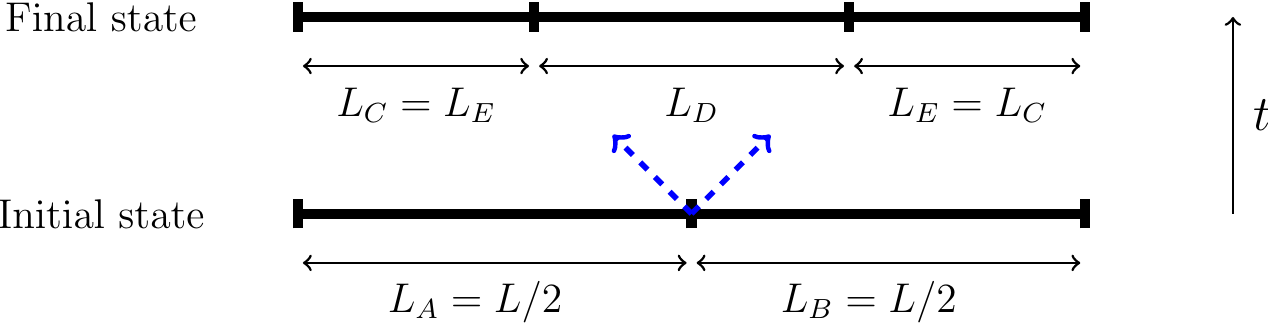}
\end{center}
\caption{The system is initially prepared in the ground state $|A\otimes B\rangle$ of $H_A+H_B$, then at time $t>0$ it evolves with the total Hamiltonian
 $H = H_A+H_B+H_{AB}^{\rm I}$. Finally it is compared with the ground state $|C\otimes D\otimes E\rangle$ of $H_C+H_D+H_E$. Blue dashed arrows
 are a graphical representation of the quasiparticles emitted after the quench.}
\label{fig:quenchline2}
\end{figure}

\begin{figure}[ht]
 \begin{center}
  \includegraphics[width=7.5cm]{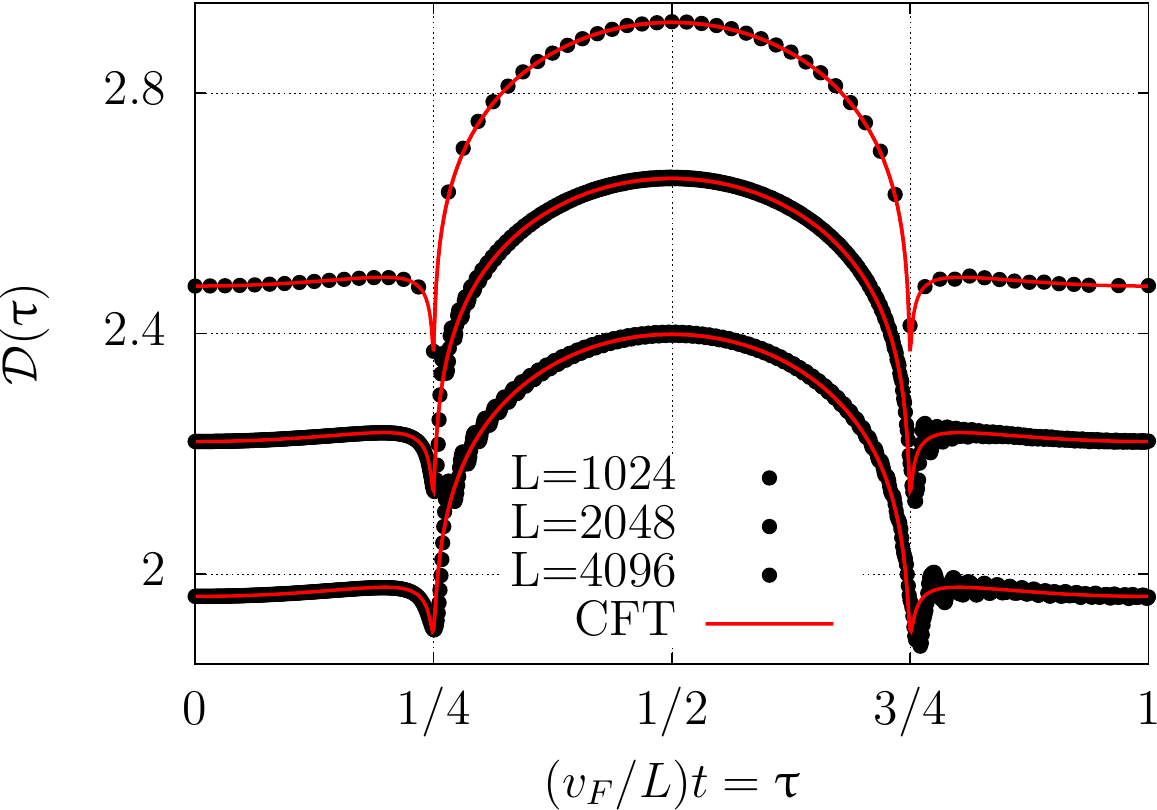}
  \includegraphics[width=7.5cm]{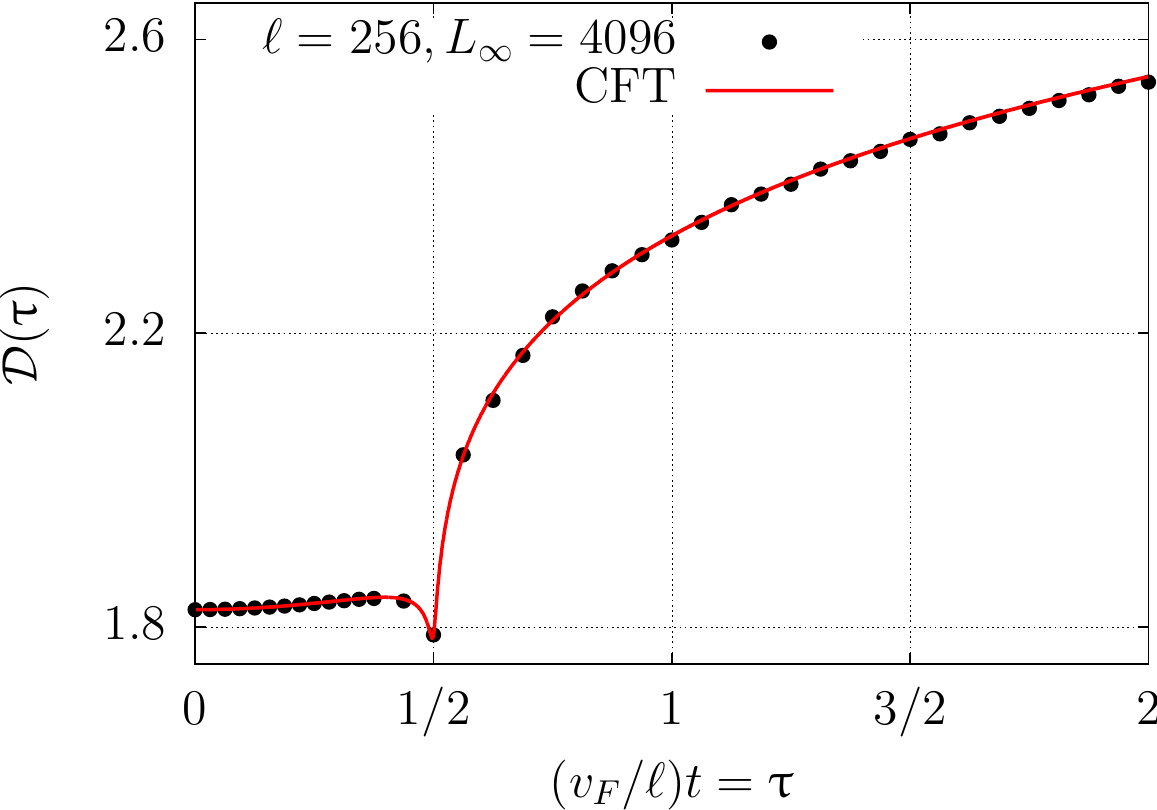}
 \end{center}
\caption{Left panel: symmetric detector. Black dots are the numerical points for system sizes $L=1024,2048,4096$, whereas red curves are the CFT prediction of
 Eq.~\ref{eq:detector_sym}. Right panel: infinite detector for $\ell=256$ and $L=L_{\infty}=4096 \gg \ell$. Black dots are the numerical points, red curve is
the prediction of Eq.~\ref{eq:detector_inf}.}
\label{fig:detector}
\end{figure}

To further probe the light-cone effects, and confirm the appearance of the plateaus as a generic phenomenon, we introduce here a variant of the (logarithmic) Loschmidt echo. 
We call it ``detector'' for reasons which will become transparent later. It is defined as
\begin{equation}
 \mathcal{D}(t)=-\log \left|\langle A \otimes B |e^{iHt}|C\otimes D\otimes E\rangle\right|^2
\end{equation}
The difference is that the wave function after a time $t$ $|\psi(t)\rangle=e^{iHt}|A\otimes B\rangle$
 is not compared with the initial state, but with another tensor product of $3$ ground-states $|C\rangle$, $|D\rangle$, and
 $|E\rangle$, see Fig.~\ref{fig:quenchline2}.
 For simplicity we only consider the case $L_A=L_B=L/2$ and $L_C=L_E$ here, so that the only remaining free length in
 the problem is the middle one $L_D$.
Now what should be the behaviour of such a quantity? Obviously the quench at $t=0$ is the same, so that there are again quasi-particles emitted
 at the interface between subsystem $A$ and $B$. They propagate at speed $v_F$, but from the point of view of the final state $|C\otimes D\otimes E\rangle$,
 they are at the beginning in the same subsystem, namely $D$. This situation is particularly interesting, because it means that the detector
 $\mathcal{D}(t)$ should start with a plateau, up to a time $t=L_D/2$, when the quasi particles emitted to the left and to the right
 simultaneously get out of $D$. In this sense, the final state is able to ``detect'' the quasiparticles. One other advantage is
 that $\mathcal{D}(t)$ can also be computed in CFT, using the same techniques as for the Loschmidt echo. The technical details are presented in
 \ref{sec:detectors_cft}. We present here only the results, focusing on two geometries:
\begin{itemize}
 \item The ``symmetric detector'' $L_D=L/2$ (and therefore $L_C=L_E=L/4$). In this case the enhanced symmetry even allows a full explicit solution
 (we set $v_F=1$):
\begin{equation}\label{eq:detector_sym}
 \mathcal{D}(t)=\frac{c}{4}\left[
\frac{1}{2}\log \left|2\cos 2\pi t/L\right|+2\log \left|1-e^{i\pi t/L}\sqrt{2\cos 2\pi t/L}\right|\right]\quad, 0\leq t \leq L/2\\
\end{equation}
This happens to be the only geometry in which the plateau can be obtained in explicit form.
$\mathcal{D}$ also satisfies $\mathcal{D}(L-t)=\mathcal{D}(t)$, $\mathcal{D}(t+L)=\mathcal{D}(t)$ and is quadratic at small times 
$\mathcal{D}(t)\propto t^2$, similar to the Loschmidt echo at long times for the semi-infinite geometry.

 \item The ``infinite detector'' $L_D=\ell \ll L$. The solution is this time not fully explicit, but $\mathcal{D}(t)$ 
can nevertheless be expressed using the Lambert
 $W$ function, solution in the complex plane of the implicit equation $w e^w=z$ :
\begin{equation}\label{eq:detector_inf}
 \mathcal{D}(t)=\frac{c}{8}\log \big|a(t)[a(t)-1] \big|\quad,\quad a(t)=W\left(e^{i\pi 2t/\ell-1}\right)+1
\end{equation}
As for the Loschmidt echo in the semi-infinite geometry, several interesting features can be extracted from this implicit solution.
 At small times $\mathcal{D}(t)$ behaves as $\propto t^2$, and the plateau lasts up to $t=\ell/2$, consistent with the quasiparticle
 interpretation. After this, the quasiparticles (and $\mathcal{D}(t)$) do not feel anymore the subtleties of the detector geometry,
 and we get\footnote{
The solution $a(t)=-\alpha(t)+i\beta(t)$ admits the following asymptotic expansion for $t\gg \ell$: $\alpha(t)\sim \log (\pi t/\ell)$
 and $\beta(t)\sim \pi t/\ell$.
} for $t/\ell \gg 1$:
\begin{equation}
 \mathcal{D}(t) \sim \frac{c}{4} \log t.
\end{equation}
We recover the result of the infinite geometry. 
\end{itemize}
\paragraph{}
We also performed numerical tests of the formula Eqs.~\ref{eq:detector_sym} and \ref{eq:detector_inf} in the XX chain
 and for very large system sizes. The numerical results in Fig.~\ref{fig:detector} are in excellent agreement with
 our analytical calculations.

%%%%%%%%%%%%%%%%%%%%%%%%%%%%%%%%%%%%%%%%%%%%%%%%%
%%%%%%%%%%%%%%%%%%%%%%%%%%%%%%%%%%%%%%%%%%%%%%%%%
\section{Conclusion}
%%%%%%%%%%%%%%%%%%%%%%%%%%%%%%%%%%%%%%%%%%%%%%%%%
%%%%%%%%%%%%%%%%%%%%%%%%%%%%%%%%%%%%%%%%%%%%%%%%%
In this paper, we study the time evolution of a one-dimensional quantum critical system after
a local quench. This ``{\it cut and glue}'' quench is particular, in the sense that it corresponds
to a single conformal boundary condition in the scaling limit. This is clearly restrictive,
and the problem of quenches corresponding to different conformal boundary conditions
(and more generically to the possible RG flows between those) deserves some further investigation.
For such a quench, we have argued that the evolution of the system is periodic,
with period $2L/v_F$ if $L$ is the size of the system and $v_F$ is the Fermi velocity.
This is true only in the continuum limit, but we observe that it is a very good
approximation for large finite-size systems, which typically exhibit a number oscillations just after the quench. For very long times however ($t \gg L/v_F$), the finite-size
effects spoil the periodicity of the system and the CFT description is not accurate
anymore.

The restriction to this particular type of quench also allows us to get simple universal results which complete the ones obtained over the last years by Calabrese and Cardy \cite{CalabreseCardy2004,CalabreseCardy2007} and Eisler and Peschel \cite{EislerPeschel2007}. Moreover, we aimed to give some further evidence
that quantities defined as simple \textit{overlaps} can mimic the behaviour of the EE \cite{DS}. Our motivation comes partly from the fact that such overlaps should be, in principle, much easier to measure in real experiments than the EE or R\'enyi entropies. Indeed, such overlaps are nothing but the {\it probability} of observing the system in its ground
state after a quench $H \rightarrow H'$. Although Cardy \cite{Cardyexp2011} recently proposed a type
of local quench that would lead to the measure of the R\'enyi entropies, it seems to
us that the ``cut and glue'' quench is still more likely to lead to a
realistic implementation. In this paper we thus focus on overlaps, and
more precisely on the (logarithmic) Loschmidt echo (which is a time-dependent 
version of the LBF introduced in \cite{DS}), and compare its
behaviour with the one of the EE. From a CFT perspective, both quantities can be computed
through analytic continuation $\tau \rightarrow v_F t+i \epsilon$, where $\epsilon$ is an UV
cutoff. However, for general geometries, it is not possible to perform this
analytic continuation to compute the EE (because of the absence of an analytic
expression for the one-point function in generic cases). On the contrary, we find
that for the LLE, a semi-analytic treatment involving a numerical solution gives
very accurate results.

We provide numerical checks of our analytic calculations in free fermion systems. We
observe light-cone effects, both for the EE and for the LLE. These effects, however,
are not quite the same for the two quantities. For a ``cut and glue'' quench
$H_A+H_B \rightarrow H = H_A +H_B +H^I_{AB}$ ($L_A<L_B$), within the first
oscillation of the system $0<t<2L/v_F$, one can distinguish two regions. The ``CFT region''
corresponds to $t < 2 L_A/v_F$ or $t > 2 L_B/v_F$, where we observe that 
both the EE and the
LLE are very well approximated by the CFT calculation. Then the ``decay region'' or
``plateau region'' ($2L_A/v_F<t<2L_B/v_F$), where the discrepancy between the behaviour
of the EE and the LLE is more important. The EE decays slowly, as observed first
by Eisler and Peschel \cite{EislerPeschel2007} in a semi-infinite system, who
interpreted this decay as an effect due to slow particles. The LLE does not
exhibit such a decay, but rather stays almost constant, leading to
pictures with flat plateaus. For large systems these plateaus are
very well approximated by the CFT
result, which predicts a behaviour that is symmetric under $t \mapsto 2L/v_F-t$.
The EE is more sensitive to non-universal
effects than the LLE for this ``cut and glue'' quench. We observe,
however, that the Loschmidt echo depends on the filling fraction of the
fermion systems which we investigate in the $c=1$ case. The CFT
results we have obtained in this paper for the LLE turn out to be correct
only at half-filling, and these must be modified for other
filling fractions, as discussed in section \ref{sec:filling} (the same phenomenon
appeared already for the LBF).

Finally, it would be interesting to have checks of all the CFT results ---which
have accumulated over the past years--- coming from other fields and using
different techniques such as integrable
systems \cite{Caux1,Caux2,Caux3} or numerical calculations in interacting systems.
Another important direction is to study how these CFT results can be
related to more general types of quenches, where the parameters in the Hamiltonian are not turned
on/off instantaneously, but rather continuously on a short time scale.
It is known that the behaviour of the system can be affected in different ways
at short time (see for example \cite{deGrandi2010}). CFT techniques lead
to results that are valid only on longer time scales. How to relate these
different regimes is an interesting open question. More generally, 
it is not clear to us what are the most relevant physical aspects
which are missed in these CFT models, and how 
one could make the latter more realistic. This seems to be of crucial importance
if one believes that such results are to be compared
with the ones of an actual experiment.

\ack
We wish to thank Jesper Jacobsen for early collaboration on this topic. We also thank
Roberto Bondesan, Gr\'egoire Misguich, Hubert Saleur and Francis Song
for several insightful discussions. We thank Pasquale Calabrese and Vincent Pasquier for very useful suggestions, for reading carefully this paper, and
 for their constant encouragements to publish these results. JD acknowledges
support from Yale University through a Yale Postdoctoral Prize Fellowship.

\begin{appendix}
\section[ \hspace{2cm}Loschmidt echo and free fermions]{Loschmidt echo and free fermions}
\subsection[\hspace{2cm}Diagonalization]{Diagonalization}
The general method to compute of the EE
is known since Peschel ~\cite{Peschel04}. For the type of local quenches we are interested in, we refer to Refs.\cite{EislerPeschel2007,EislerPeschel2008},
 where the procedure is explained in details. Calculating the EE usually amounts to diagonalizing a $2\L_A \times 2L_A$ matrix. 
\paragraph{}
We give here some details about the numerical computation of the Loschmidt echo in free fermionic systems. 
We consider the $XY$ chain in transverse field
\begin{equation}
 H_{A\cup B}=-\frac{1}{2}\sum_{j=1}^{L-1}\left( \frac{1+r}{2}\sigma_j^x\sigma_{j+1}^x +\frac{1-r}{2}\sigma_j^y \sigma_{j+1}^y\right) 
-\frac{h}{2}\sum_{j=1}^L \sigma_j^z
\end{equation}
A Jordan-Wigner transformation
\begin{eqnarray}
 \sigma_j^x+i\sigma_j^y&=&2c_j^\dag \exp \left(\sum_{\ell=1}^{j-1}c_\ell^\dag c_\ell\right)\\
\sigma_j^z &=& 2c_j^\dag c_j -1
\end{eqnarray}
then allows to express the Hamiltonian as a quadratic form in the fermionic operators
\begin{equation}
 H_{A\cup B}=-\frac{1}{2}\sum_{j=1}^{L-1}\left[\left(c_j^\dag c_{j+1}+c_{j+1}^\dag c_j\right)+r\left(c_j^\dag c_{j+1}^\dag + c_{j+1}c_j\right)\right]
 -\frac{h}{2}\sum_{j=1}^L \left(2c_j^\dag c_j-1 \right)
\end{equation}
which can be diagonalized using a Bogoliubov transformation:
\begin{eqnarray}\label{eq:basis_chgt}
 f_m^\dag&=& \sum_{j=1}^L u_{jm} c_j^\dag +v_{jm} c_j\qquad,\qquad f_m=\left(f_m^\dag\right)^\dag
\end{eqnarray}
The Hamiltonian is then rewritten as
\begin{equation}
 H_{A\cup B}=E_0+\sum_{m} \epsilon_m f_m^\dag f_m,
\end{equation}
where we can choose all the one-particle eigenenergies to be positive $\epsilon_m>0 \;\forall m$. $E_0$
 is therefore the ground-state energy. 
Denoting by $|0\rangle$ the fermion vacuum ($c_j|0\rangle=0$), the ground-state of the total chain can then be expressed as
\begin{equation}
 |A\cup B\rangle=f_1 f_2 \ldots f_L |0\rangle.
\end{equation}
One can also do the same for $H_{A\otimes B}$. Denoting by $\{d_m,d_m^\dag\}$ the set of fermions which diagonalize
 the Hamiltonian,
\begin{equation}
 d_m^\dag=x_{jm}c_j^\dag +y _{jm}c_j
\end{equation}
 the ground-state is
\begin{equation}
 |A\otimes B\rangle=d_1 d_2 \ldots d_L|0\rangle.
\end{equation}
\subsection[\hspace{2cm}Loschmidt echo and bipartite fidelity]{Loschmidt echo and bipartite fidelity}
Recall that the Loschmidt echo is given by 
\begin{equation}
 \mathcal{L}(t)=\left|\langle A\otimes B|\exp\left(itH_{A\cup B}\right)|A\otimes B\rangle\right|^2
\end{equation}
To compute it, it suffice to notice that the state
\begin{equation}
 |\Omega\rangle= d_1 \ldots d_L |A\cup B\rangle
\end{equation}
is an eigenstate of $H_{A\otimes B}$ with the minimum eigenenergy $E_0$. Since the ground state of such an Hamiltonian is unique (Perron Frobenius theorem),
 $|\Omega\rangle$ is actually proportional to $|A\otimes B\rangle$. We therefore have
\begin{equation}
 \mathcal{L}(t)=\frac
{\left|\langle A\cup B|d_1^\dag \ldots d_L^\dag e^{itH_{A\cup B}}d_1\ldots d_L|A\cup B\rangle\right|^2}
{\left|\langle A\cup B |d_1^\dag \ldots d_L^\dag d_1 \ldots d_L|A\cup B\rangle\right|^2}
\end{equation}
The denominator is
nothing but $|\langle A\cup B|A\otimes B\rangle|^4$, the (squared) bipartite fidelity\cite{DS}, and ensures
 $\mathcal{L}(t=0)=1$.
 The numerator $\mathcal{N}(t)$ can be
 expressed using the fermions operators
\begin{equation}
 \mathcal{N}(t)=
\big|\langle d_1^\dag \ldots d_L^\dag
d_1(t)\ldots d_L(t)\rangle\big|^2
\end{equation}
where $d_m(t)=e^{itH_{A\cup B}}d_me^{-itH_{A\cup B}}$ 
 and we have used the fact that $e^{itH_{A\cup B}}|A\cup B\rangle$ contributes just to a phase. This correlator may be reduced
 to a Pfaffian using Wick's theorem. Using the relation ${\rm Pf}^2=\det$, 
\begin{equation}
 \mathcal{N}(t)=\left|\det \left(
\begin{array}{cc}
 M&N^{(t)}\\
-N^{(t)}&-M
\end{array}
\right)\right|,
\end{equation}
where $M$ is antisymmetric, while $N^{(t)}$ is symmetric. The matrix elements are given by
\begin{eqnarray}
 M_{j\ell}&=&\langle d_j^\dag d_\ell^\dag \rangle=-\langle d_j d_\ell \rangle=-\langle d_j(t) d_\ell(t) \rangle\\
N_{j\ell}^{(t)}&=&\langle d_j^\dag d_\ell(t)\rangle
\end{eqnarray}
In the determinant, adding the second column to the first and then the first row to the second we find
\begin{equation}
 \mathcal{N}(t)=\left|\det \left(
\begin{array}{cc}
 N^{(t)}+M&N^{(t)}\\
0&N^{(t)}-M
\end{array}
\right)\right|=
\left|\det (N^{(t)}+M)\right|^2
\end{equation}
\subsection[\hspace{2cm}Bipartite fidelity]{Bipartite fidelity}
The bipartite fidelity\cite{DS} is the (squared) overlap $|\langle A\cup B|A\otimes B\rangle|^2$, and is a direct by-product
 of the preceding calculation. It can therefore be expressed as $L\times L$ determinant:
\begin{equation}
 |\langle A \cup B|A\otimes B\rangle|^2=\left|\det_{j,\ell} \left(\big\langle d_j^\dag (d_\ell^\dag +d_\ell)\big\rangle\right)\right|.
\end{equation}
 This result was actually already used in Ref.~\cite{DS}.
%\subsection[\hspace{2cm}A simple example: XX chain]{A simple example: XX chain}
\section[ \hspace{2cm} Details on the detector geometries]{Details on the detector geometries}
\label{sec:detectors_cft}
\subsection[ \hspace{2cm} The ``symmetric detector'']{The ``symmetric detector''}
\begin{figure}[ht]
\begin{center}
\includegraphics{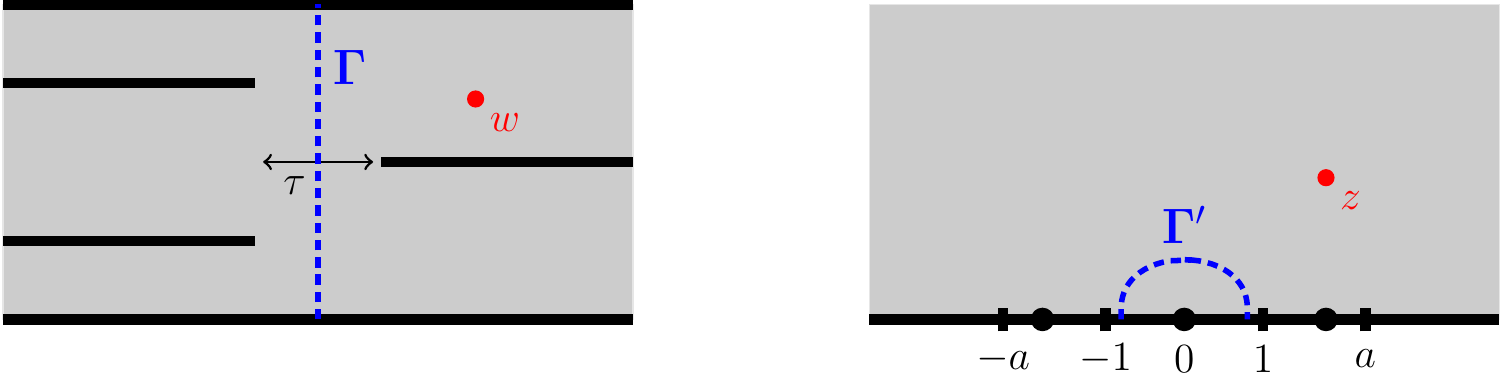}
\end{center}
\caption{``Detector'' geometry (left) and upper half plane $\mathbb{H}$ (right). 
The conformal mapping from the latter to the former is given by Eq.~\ref{eq:conf_detector}.}
\label{fig:sym_detector}
\end{figure}

This turns out to be simplest detector geometry. The conformal transformation which maps the upper half-plane onto
 a strip with $3$ slits of Fig.~\ref{fig:sym_detector} is:
\begin{equation}\label{eq:conf_detector}
 w(z)=\frac{L}{4\pi}\left[ \log (1-z^2)-2 \log (1-a^2 z^2) \right]
\end{equation}
This mapping has singularities at $z=\pm 1$, $z=\pm a$ and extrema at $z=0$, $z=\pm\sqrt{2-1/a^2}$. The distance $\tau$
 between the two slits is
\begin{equation}\label{eq:sym_detector_tau}
 \tau(a)=\frac{L}{4\pi} \log \left(4 a^2(a^2-1)\right)
\end{equation}
The crucial point is that this bisquare relation can easily be inverted. 
We then use an infinitesimal transformation $w\mapsto (1+\delta \tau/\tau)w$ at the right of path $\mathbf{\Gamma}$,
 $w\mapsto w$ at the left. Denoting as before by $f(z)=S(w,z)(dw/dz)^{-1}$, the variation of the free energy is
\begin{equation}
 \frac{\delta \mathcal{D}}{\delta \tau}=\frac{c}{6\pi}\int_{\mathbf{\Gamma'}} f(z)\,dz
\end{equation}
where $\mathbf{\Gamma'}$ is the image of $\mathbf{\Gamma}$ under the inverse map $w\mapsto z$ (see Fig.~\ref{fig:sym_detector}).
 The curve $\mathbf{\Gamma'}$ can be closed onto a contour using its mirror image along the real axis, and the calculation boils down to a residue.
\begin{equation}
 \frac{\delta \mathcal{D}}{\delta \tau}=\frac{c}{6} \times{\rm Res}[f(z);z=0],
\end{equation}
so that 
\begin{equation}
\mathcal{D}(a)=\frac{c}{8}\log \left|\frac{(a^2-1)(2a^2-1)}{a^2}\right|
\end{equation}
Inverting Eq.~\ref{eq:sym_detector_tau} with $\tau=it$ we finally get
\begin{equation}
 \mathcal{D}(t)=\frac{c}{4}\left[
\frac{1}{2}\log \left|2\cos 2\pi t/L\right|+2\log \left|1-e^{i\pi t/L}\sqrt{2\cos 2\pi t/L}\right|\right]\quad, 0\leq t \leq L/2\\
\end{equation}
and $\mathcal{D}(L-t)=\mathcal{D}(t+L)=\mathcal{D}(t)$. In this equation the square root has to be understood as $\sqrt{\alpha}=i\sqrt{-\alpha}$ for real
 negative $\alpha$.
\subsection[ \hspace{2cm} The ``infinite detector'']{The ``infinite detector''}
Here the required conformal transformation reads
\begin{equation}\label{eq:conf_inf_detector}
 w(z)=\frac{\ell}{2\pi}\left[\frac{a-1}{1-z^2}-\log \left(z-1\right)-\log (z+1)\right],
\end{equation}
\begin{figure}[ht]
\begin{center}
\includegraphics{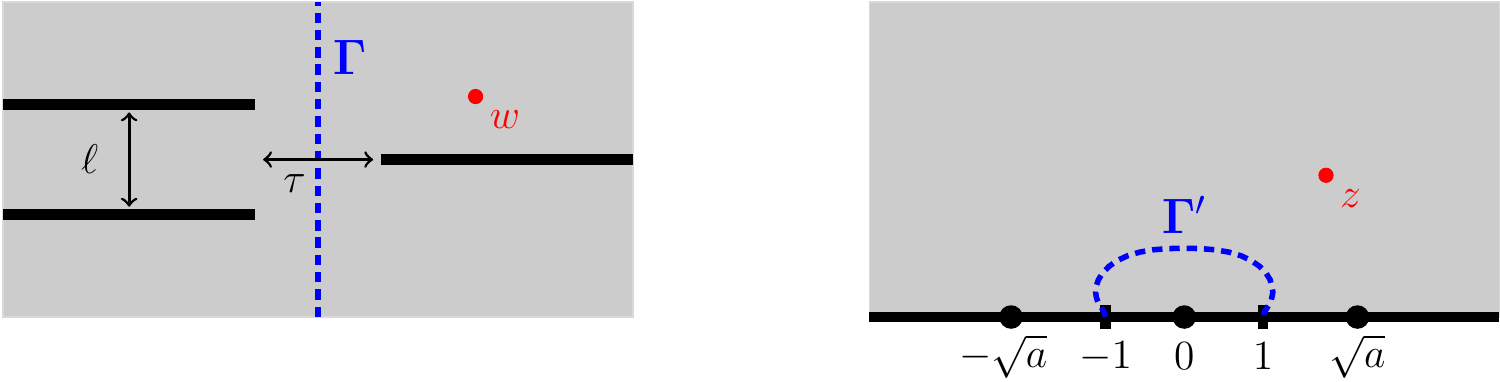}
\end{center}
\caption{``Infinite Detector'' geometry (left) and upper half plane $\mathbb{H}$ (right). 
The conformal mapping from the latter to the former is given by Eq.~\ref{eq:conf_inf_detector}}
\end{figure}
has singularities at $z=\pm 1$ and extrema at $z=0$, $z=\pm \sqrt{a}$. The distance between the two slits is
\begin{equation}\label{eq:detector_inf_tau}
 \tau(a)=\frac{\ell}{2\pi}\left[a+\log (a-1)\right]
\end{equation}
The detector in imaginary time is solution of the equation
\begin{equation}
 \frac{\delta \mathcal{D}}{\delta \tau}=\frac{c}{6}\times {\rm Res}[f(z);z=0],
\end{equation}
which can easily be integrated
\begin{equation}
 \mathcal{D}(a)=\frac{c}{8}\log \left|a(a-1)\right|
\end{equation}
Contrary to the previous case, Eq.~\ref{eq:detector_inf_tau} cannot be explicitly inverted. The solution can nevertheless
 be expressed as
\begin{equation}
 a(t)=1+W(e^{i\pi 2t/\ell-1})
\end{equation}
where $W$ is the (multivalued) Lambert $W$ function, solution of the implicit equation $w e^w=z$. This is especially useful when
 trying to evaluate the Detector, because $W$ is part of the toolbox of usual symbolic computer programs. 
\section[ \hspace{2cm} Intriguing trajectories in the complex plane]{Intriguing trajectories in the complex plane}
\label{sec:trajectories}
\begin{figure}[ht]
\begin{center}
\includegraphics[width=5cm]{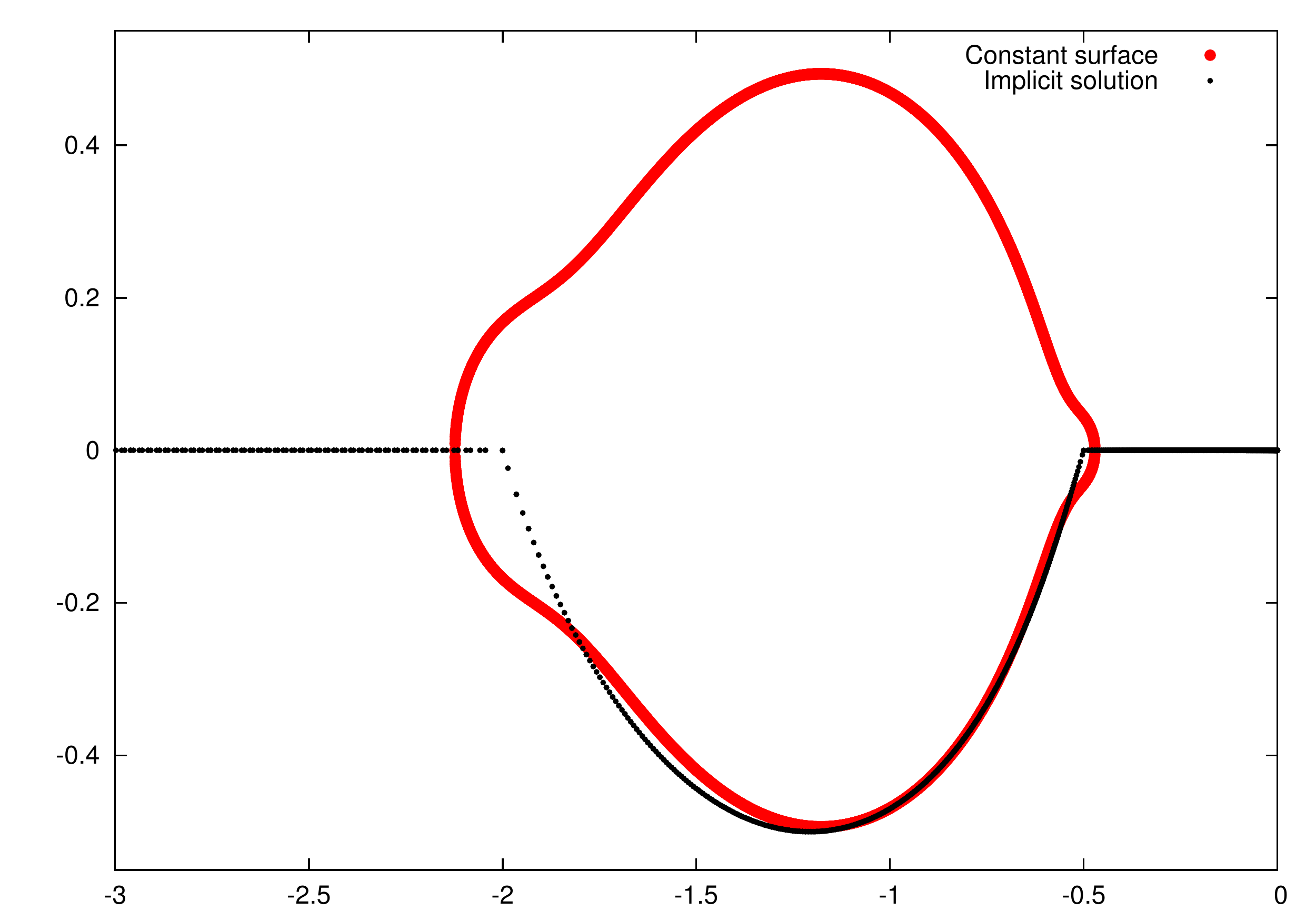}
\includegraphics[width=5cm]{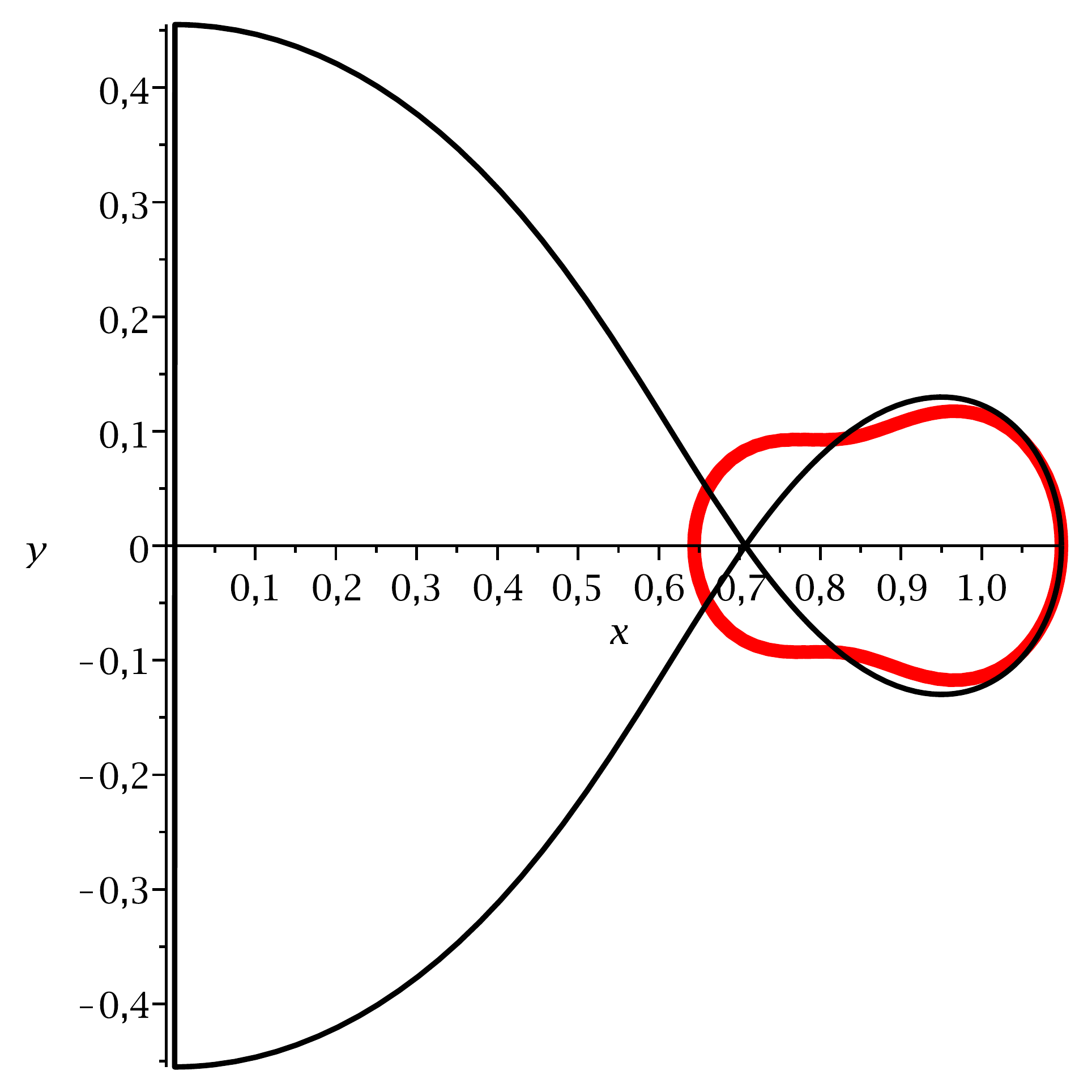}
\includegraphics[width=5cm]{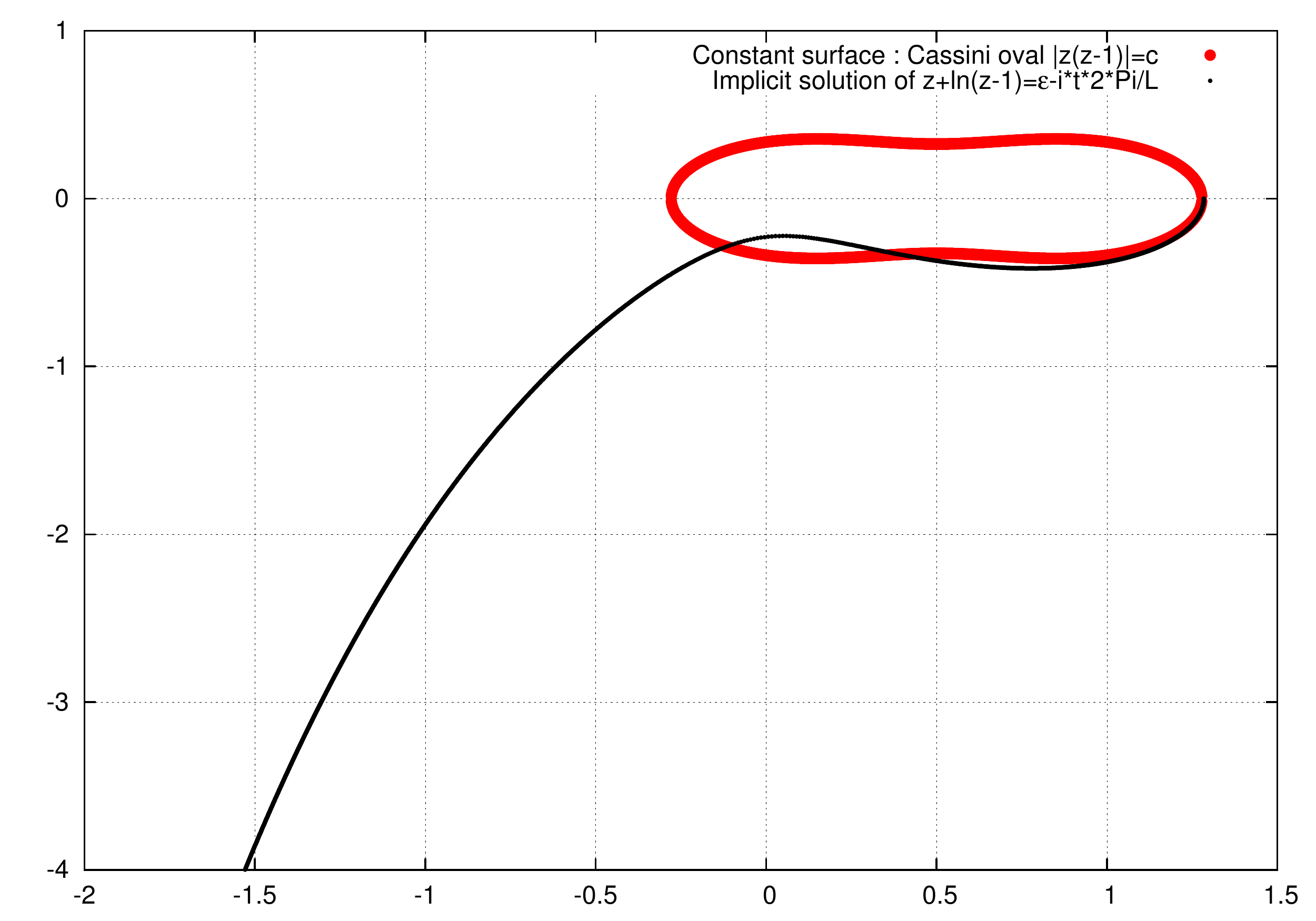}
\end{center}
\caption{Left panel: Loschmidt echo for $x=\ell/L=1/3$. Middle panel: ``symmetric detector''. Right panel: ``infinite detector''.
 Black curve is $a(t)$. The fat red curve is 
Red curve is the iso Loschmidt echo $\mathcal{L}(a)=\mathcal{L}\left[a(t=L)\right]$ for the left panel,
 and the iso Detector $\mathcal{D}(a)=\mathcal{D}\left[a(t=0)\right]$ for the two other.
 The plateau correspond to times when the black curve $a(t)$ almost follows the red one.}
\label{fig:a(t)}
\end{figure}

In all cases we encountered in this paper, the Loschmidt echo (or the detector) could be expressed as 
\begin{equation}
 -\log \mathcal{L}(t)=-\log \mathcal{L}\left[a(t)\right],
\end{equation}
where except in favourable cases $a(t)$ is the solution of some implicit equation in the complex plane. The goal of this section
 is to make this solution more intuitive, by simply showing this parametric function in the complex plane, especially at times corresponding
 to a plateau. 
\paragraph{}
We focus on three geometries : the asymmetric Loschmidt echo for $x=\ell/L=1/3$, the ``symmetric detector'', and finally 
 the ``infinite detector''. The results are summarized in Fig.~\ref{fig:a(t)}, where $a(t)$ is shown for these three geometries.
For the Loschmidt echo we also show the curve solution of $\mathcal{L}(a)=\mathcal{L}\left[a(t=L)\right]$, namely the set
 of $a$ which would give the same values as the observed one in the middle of the plateau at $t=L$. 
This is the red curve in the left panel of Fig.~\ref{fig:a(t)}.
 Since the plateau region is close to perfect, one expects the solution $a(t)$ to be
 very close to this curve at times $2L_A\leq t \leq 2L_B$, and this is indeed what happens. 
We also performed the same analysis for the detector, now showing the curves solution of
 $\mathcal{D}(a)=\mathcal{D}\left[(a(t=0))\right]$, since the plateau begins at time $t=0$ in this case.
 We also observe the same phenomenon. As an aside, we note that the red ``iso-detector'' curve
 is a well-known Cassini oval.

\end{appendix}

\end{document}